\documentclass{jpp}
\usepackage{graphicx}
\usepackage{tensor}
\usepackage{bm}

\usepackage[utf8]{inputenc}
\usepackage[T1]{fontenc}
\usepackage{amsmath}
\DeclareMathOperator{\sech}{sech}

\shorttitle{Parametric Instability and Inverse Cascade}
\shortauthor{B. D. G. Chandran}

\title{Parametric Instability, Inverse Cascade, and the $1/f$ Range of
  Solar-Wind Turbulence}

\author{Benjamin D. G. Chandran \corresp{\email{benjamin.chandran@unh.edu}}}

\affiliation{Department of Physics, University of New Hampshire, Durham, New Hampshire 03824,
  USA}

\begin{document}

\maketitle

\begin{abstract}
  In this paper, weak turbulence theory is used to investigate the nonlinear
  evolution of the parametric instability in 3D low-$\beta$ plasmas at
  wavelengths much greater than the ion inertial length under the
  assumption that slow magnetosonic waves are strongly damped.  It is
  shown analytically that the parametric instability leads to an
  inverse cascade of Alfv\'en wave quanta, and several exact solutions
  to the wave kinetic equations are presented.  The main results of
  the paper concern the parametric decay of Alfv\'en waves that
  initially satisfy $e^+ \gg e^-$, where $e^+$ and $e^-$ are the
  frequency ($f$) spectra of Alfv\'en waves propagating in opposite
  directions along the magnetic field lines.  If $e^+$ initially has a
  peak frequency~$f_0$ (at which $f e^+$ is maximized) and an
  ``infrared'' scaling~$f^p$ at smaller~$f$ with $-1 < p < 1$, then
  $e^+$ acquires an $f^{-1}$ scaling throughout a range of frequencies
  that spreads out in both directions from~$f_0$. At the same time,
  $e^-$ acquires an $f^{-2}$ scaling within this same frequency range.
  If the plasma parameters and infrared $e^+$ spectrum are chosen to
  match conditions in the fast solar wind at a heliocentric distance
  of 0.3 astronomical units (AU), then the nonlinear evolution of the
  parametric instability leads to an $e^+$ spectrum that matches
  fast-wind measurements from the {\em Helios} spacecraft at 0.3~AU,
  including the observed $f^{-1}$ scaling at
  $f \gtrsim 3 \times 10^{-4} \mbox{ Hz}$. The results of this paper
  suggest that the $f^{-1}$ spectrum seen by {\em Helios} in the fast
  solar wind at $f \gtrsim 3\times 10^{-4} \mbox{ Hz}$ is produced in
  situ by parametric decay and that the $f^{-1}$ range
  of $e^+$ extends over an increasingly narrow range of frequencies as
  $r$ decreases below 0.3~AU. This prediction will be tested by
  measurements from the {\em Parker Solar Probe}.
\end{abstract}

\section{Introduction}
\label{sec:intro} 

The origin of the solar wind is a long-standing problem
\citep{parker58} that continues to receive considerable
attention. A leading model for the origin of the fast solar wind
appeals to Alfv\'en waves (AWs) that are launched by photospheric
motions. As these AWs propagate away from the Sun, they undergo
partial reflection due to the radial variation of the Alfv\'en speed
\citep{heinemann80}. Nonlinear interactions between
counter-propagating AWs then cause AW energy to cascade to small
scales and dissipate, heating the plasma
\citep{velli89,zhou89,cranmer05,verdini12,perez13,vanballegooijen17}.
This heating increases the plasma pressure, which, in conjunction with
the wave pressure, accelerates the plasma to high speeds
\citep{suzuki05,cranmer07,verdini10,chandran11,vanderholst14}.

Although non-compressive AWs are the primary mechanism for energizing
the solar wind in this model, a number of considerations indicate that
compressive fluctuations have a significant impact on the dynamics of
turbulence in the corona and solar wind. Observations of the tail of
Comet-Lovejoy reveal that the background plasma density~$\rho_0$ at
$r= 1.2 R_{\odot}$ (where $R_{\odot}$ is the radius of the Sun) varies
by a factor of~$\sim 6$ over distances of a few thousand~km measured
perpendicular to the background magnetic field~$\bm{B}_0$
\citep{raymond14}. These density variations (denoted $\delta \rho$) lead to
phase mixing of AWs, which transports AW energy to smaller scales
measured perpendicular to~$\bm{B}_0$ \citep{heyvaerts83}. Farther from
the Sun, where $\delta \rho/\rho_0$ is significantly smaller than
$ |\delta \bm{B}|/B_0$ \citep{tumarsch95,hollweg10}, AWs still couple
to slow magnetosonic waves (``slow waves'') through the parametric
instability, in which outward-propagating AWs decay into
outward-propagating slow waves and inward-propagating AWs\footnote{The
  terms outward-propagating and inward-propagating refer to the
  propagation direction in the plasma rest frame. Beyond
  the Alfv\'en critical point, all AWs
  propagate outward in the rest frame of the Sun.}
\citep{galeev63,sagdeev69,goldstein78,spangler86,spangler89,spangler90,hollweg94,dorfman16}.
This instability and its nonlinear evolution are the focus of the
present work.

A number of studies have investigated the parametric instability in
the solar wind within the framework of magnetohydrodynamics (MHD)
\citep[e.g.,][]{malara00,delzanna01a,shi17}, while others have gone
beyond MHD to account for temperature anisotropy \citep{tenerani17} or
kinetic effects such as the Landau damping of slow waves
\citep[e.g.][]{inhester90,vasquez95,araneda08,maneva13}.
\cite{cohen74}, for example, derived the growth rate of the parametric
instability in the presence of strong slow-wave damping and randomly
phased, parallel-propagating AWs. \cite{terasawa86} carried out
1D hybrid simulations and found that Landau damping reduces the growth
rate of the parametric instability and that the parametric instability
leads to an inverse cascade of AWs to smaller frequencies.

In this paper, weak turbulence theory is used to investigate the
nonlinear evolution of the parametric instability assuming a randomly
phased collection of AWs at wavelengths much greater than the proton
inertial length~$d_{\rm i}$ in a low-$\beta$ plasma, where $\beta$ is
the ratio of plasma pressure to magnetic pressure. The fluctuating
fields are taken to depend on all three spatial coordinates, but the
wave kinetic equations are integrated over the perpendicular (to
$\bm{B}_0$) wave-vector components, yielding equations for the 1D
power spectra that depend only on the parallel wavenumber and
time. The starting point of the analysis is the theory of weak
compressible MHD turbulence.  Collisionless
damping of slow waves is incorporated in a very approximate manner
analogous to the approach of \cite{cohen74}, by dropping terms
containing the slow-wave energy density in the wave kinetic equations
that describe the evolution of the AW power spectra.

The remainder of the paper is organized as
follows. Section~\ref{sec:WKE} reviews results from the theory of weak
compressible MHD turbulence, and Section~\ref{sec:linear} uses
the weak-turbulence wave kinetic equations
to recover the results of \cite{cohen74} in the linear
regime. Section~\ref{sec:inverse} shows
how the wave kinetic equations imply that AW quanta undergo an inverse
cascade towards smaller parallel wavenumbers, and
Section~\ref{sec:exact} presents several exact solutions
to the wave kinetic equations. The main results of the paper appear
in Section~\ref{sec:SW}, which uses a numerical solution and an
approximate analytic solution to the wave kinetic equations to
investigate the parametric decay of an initial population of randomly
phased AWs propagating in the same direction with negligible initial
power in counter-propagating AWs. The numerical results are compared
with observations from the {\em Helios} spacecraft at a heliocentric
distance of 0.3~AU. Section~\ref{sec:applicability} critically
revisits the main assumptions of the analysis and the relevance of the
analysis to the solar wind. Section~\ref{sec:conclusion} summarizes
the key findings of the paper, including predictions that will
be tested by NASA's {\em Parker Solar Probe}.

\section{The Wave Kinetic Equations for Alfv\'en Waves Undergoing Parametric Decay}
\label{sec:WKE} 

In weak turbulence theory, the quantity $\omega_{\rm nl}/\omega_{\rm
  linear}$ is treated as a small parameter, 
 where $\omega_{\rm nl}$ is the inverse of the timescale on which
nonlinear interactions modify the fluctuations, and $\omega_{\rm linear} $ is
the linear wave frequency. Because 
\begin{equation}
\omega_{\rm nl} \ll \omega_{\rm linear},
\label{eq:wto}
\end{equation} 
the fluctuations can be viewed as waves to a good approximation. The
governing equations lead to a hierarchy of equations for the moments
of various fluctuating quantities, in which the time derivatives of
the second moments (or second-order correlation functions) depend upon
the third moments, and the time derivatives of the third moments
depend upon the fourth moments, and so on. This system of equations is
closed via the random-phase approximation, which allows the
fourth-order correlation functions to be expressed as products of
second-order correlation functions \citep[see, e.g.,][]{galtier00}.

The strongest nonlinear interactions in weak MHD turbulence are
resonant three-wave interactions. These interactions occur when the
frequency and wavenumber of the beat wave produced by two waves is
identical to the frequency and wavenumber of some third wave, which
enables the beat wave to drive the third wave coherently in time.  If
the three waves have wavenumbers $\bm{p}$, $\bm{q}$, and $\bm{k}$ and
frequencies $\omega_p$, $\omega_q$, and $\omega_k$, respectively, then
a three-wave resonance requires that
\begin{equation}
\bm{k} = \bm{p} + \bm{q}
\label{eq:kres} 
\end{equation} 
and 
\begin{equation}
\omega_k = \omega_p + \omega_q.
\label{eq:omegares} 
\end{equation} 
An alternative interpretation of Equations~(\ref{eq:kres}) and
(\ref{eq:omegares}) arises from viewing the wave fields as a
collection of wave quanta at different wavenumbers and frequencies,
restricting the frequencies to positive values,  and
assigning a wave quantum at wavenumber~$\bm{k}$ and
frequency~$\omega_k$ the momentum~$\hbar \bm{k}$ and
energy~$\hbar \omega_k$.  Equations~(\ref{eq:kres}) and
(\ref{eq:omegares}) then correspond to the momentum-conservation
and energy-conservation relations that arise when either one wave quantum decays
into two new wave quanta or two wave quanta merge to produce a new
wave quantum.

In the parametric instability in a low-$\beta$ plasma, a parent AW (or
AW quantum) at wavenumber~$\bm{k}$ decays into a slow wave at
wavenumber~$\bm{p}$ propagating in the same direction and an AW at
wavenumber~$\bm{q}$ propagating in the opposite direction. Regardless
of the direction of the wave vector, the group velocity of an AW is
either parallel or anti-parallel to the background magnetic field
\begin{equation}
\bm{B}_0 = B_0 \bm{\hat{z}},
\label{eq:B0} 
\end{equation} 
and the same is true for slow waves when
\begin{equation}
\beta \ll 1,
\label{eq:lowbeta} 
\end{equation} 
which is henceforth assumed. At low~$\beta$ slow
waves travel along field lines at the sound speed~$c_{\rm s}$, which is roughly
$\beta^{1/2}$ times the Alfv\'en speed~$v_{\rm A}$.
Thus, regardless of the perpendicular components of~$\bm{k}$,
$\bm{p}$, and~$\bm{q}$, the frequency-matching condition
(Equation~(\ref{eq:omegares})) for the parametric instability is 
\begin{equation}
k_z v_{\rm A} = p_z c_{\rm s} - q_z v_{\rm A}.
\label{eq:omegares2} 
\end{equation} 
Combining the $z$ component of Equation~(\ref{eq:kres}) with
Equation~(\ref{eq:omegares2})  and taking $c_{\rm s} \ll v_{\rm A}$ yields
\begin{equation}
p_{\rm z} \simeq 2 k_z
\label{eq:pz1} 
\end{equation} 
and
\begin{equation}
q_z \simeq -k_z \left(1 - \frac{2 c_{\rm s}}{v_{\rm A}}\right).
\label{eq:omegares3} 
\end{equation} 
Equation~(\ref{eq:omegares3}) implies that the
frequency $|q_z v_{\rm A}|$ of the daughter AW is slightly smaller than
the frequency $|k_z v_{\rm A}|$ of the parent~AW
\citep{sagdeev69}. Thus, the energy of the daughter AW is slightly
smaller than the energy of the parent AW. This reduction in AW
energy is offset by an increase in slow-wave energy.

\cite{chandran08b} derived the wave kinetic equations for weakly
turbulent AWs, slow waves, and fast magnetosonic waves (``fast
waves'') in the low-$\beta$ limit. The resulting
equations were expanded in powers of~$\beta$, and only the first two
orders in the expansion (proportional to $\beta^{-1}$
and $\beta^0$, respectively) were retained.
Slow waves are strongly
damped in collisionless low-$\beta$ plasmas~\citep{barnes66}.  
\cite{chandran08b} neglected collisionless damping 
during the derivation of the wave kinetic
equations, but incorporated it afterward in an ad hoc manner by
assuming that the slow-wave power spectrum~$S^\pm_k$ was small and 
discarding terms~$\propto S^\pm_k$ unless they were also
proportional to~$\beta^{-1}$.\footnote{The one exception to this rule
  was that \cite{chandran08b} retained the term representing turbulent
  mixing of slow waves by AWs, since this term can dominate the
  evolution of slow waves at small~$k_z$ \citep[][]{lithwick01,schekochihin16}.}
(The $\pm$ sign in $S^\pm_k$ indicates slow waves propagating parallel
($+$) or anti-parallel ($-$) to~$\bm{B}_0$.)

In the present paper, the wave kinetic equations derived by
\cite{chandran08b} are used to investigate the nonlinear
evolution of the parametric instability. 
It is assumed that
slow-wave damping is sufficiently strong that all terms $\propto
S_k^\pm$, even those~$\propto \beta^{-1}$, can be safely discarded.
All other types of nonlinear
interactions are neglected, including resonant interactions between three AWs, phase
mixing, and
resonant interactions involving fast waves.
Given these approximations, Equation~(8) of \cite{chandran08b} becomes
\begin{equation}
\frac{\partial A^\pm_k}{\partial t} = \frac{\pi }{v_{\rm A}}
\int d^3p\,d^3q\, \delta(\bm{k} - \bm{p} - \bm{q}) \delta (q_z + k_z)
k_z^2 A^\pm_k \frac{\partial}{\partial q_z} \left(q_z A^\mp_q\right),
\label{eq:Aweak} 
\end{equation} 
where $A^+_k$ ($A^-_k$) is the 3D wavenumber spectrum of AWs
propagating parallel (anti-parallel) to~$\bm{B}_0$, $\delta(x)$ is the
Dirac delta function, and the integral over each Cartesian component
of $\bm{p}$ and $\bm{q}$ extends from $-\infty$ to $+\infty$. The 3D
AW power spectra depend upon all three wave-vector components and time. The
$\delta (\bm{k} - \bm{p} - \bm{q})$ term enforces the wavenumber-resonance
condition (Equation~(\ref{eq:kres})), and the $\delta (q_z + k_z)$ term
enforces the frequency-resonance condition
(Equation~(\ref{eq:omegares3})) to leading order in~$\beta$. The
integral over the components of~$p$ in Equation~(\ref{eq:Aweak}) can
be carried out immediately, thereby annihilating the first delta
function.  Equation~(\ref{eq:Aweak}) can be further simplified by introducing the
1D wavenumber spectra
\begin{equation}
E^\pm(k_z,t) = \int_{-\infty}^\infty dk_x \int_{-\infty}^\infty dk_y
A^\pm_k
\label{eq:defEpm} 
\end{equation} 
and integrating Equation~(\ref{eq:Aweak})  over~$k_x$ and~$k_y$, which yields
\begin{equation}
\frac{\partial E^\pm}{\partial t} = \frac{\pi }{v_{\rm A}} k_z^2 E^\pm
\frac{\partial}{\partial k_z} \left( k_z E^\mp\right).
\label{eq:dEpmdt} 
\end{equation} 
Equation~(\ref{eq:dEpmdt}) describes how the 1D (parallel) power
spectra~$E^\pm$ evolve and forms the basis for much of the
discussion to follow. Given the aforementioned assumptions, the
evolution of the 1D power spectra $E^\pm$ is not influenced by the way
that~$A^\pm$ depends on $k_x$ and~$k_y$. For future reference,
the normalization of the power spectra is such that
\begin{equation}
\int_{-\infty}^\infty dk_z E^\pm  = \frac{1}{2} \left\langle
  \left|\delta \bm{v}_{\rm AW}
\mp  \frac{\delta \bm{B}_{\rm AW}}{\sqrt{4 \pi \rho}} \right|^2\right\rangle,
\label{eq:Epmnorm} 
\end{equation} 
where $\delta \bm{v}_{\rm AW}$ and $\delta \bm{B}_{\rm AW}$ are the
velocity and magnetic-field fluctuations associated with AWs, and
$\langle \dots \rangle$ indicates an average over space and time
\citep[][]{chandran08b}.

\begin{figure}
\centerline{
\includegraphics[width=12cm]{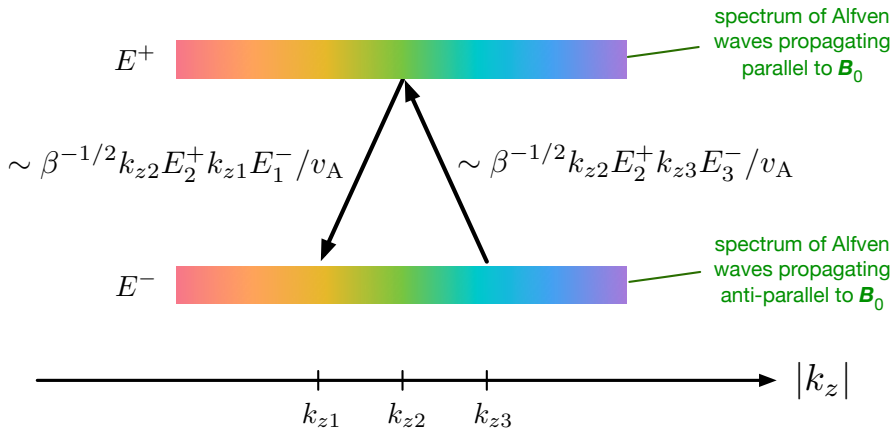}
}
\caption{Physical interpretation of the wave kinetic equation for
  parametric decay when slow waves are strongly damped (Equation~(\ref{eq:dEpmdt})). The
  mathematical expressions next to the arrows represent the
  contributions to $\partial E^+(k_{z2})/\partial t$ from the
  parametric decay of AWs at~$k_{z3}$, which acts to increase
  $E^+(k_{z2})$, and the parametric decay of AWs at~$k_{z2}$, which
  acts to decrease~$E^+(k_{z2})$. In these expressions,
  $E^+_2 = E^+(k_{z2})$, $E^-_1 = E^-(k_{z1})$,
  and~$E^-_3 = E^-(k_{z3})$. \label{fig:color_spectrum}}
\end{figure}

\subsection{Physical Interpretation of the Wave Kinetic Equation}
\label{sec:interpretation} 

Figure~\ref{fig:color_spectrum} offers a way of understanding 
Equation~(\ref{eq:dEpmdt}).  
The horizontal color bars in this figure represent the spectra of
outward-propagating and inward-propagating AWs, with red representing
longer-wavelength waves and violet representing
shorter-wavelength waves. AWs propagating in the $-\bm{B}_0$ direction
at $|k_z| = k_{z3}$ decay into
slow waves propagating anti-parallel to~$\bm{B}_0$
at $|k_z| \simeq 2 k_{z3}$ and AWs propagating parallel to~$\bm{B}_0$
at $|k_z| = k_{z2}$. AWs propagating parallel to~$\bm{B}_0$ at $|k_z|
= k_{z2}$ decay into slow waves propagating parallel to $\bm{B}_0$ at
$|k_z| \simeq 2 k_{z2}$ and AWs propagating anti-parallel to
$\bm{B}_0$
at $|k_z| = k_{z1}$.
Equation~(\ref{eq:dEpmdt})  is approximately equivalent to the
statement that the rate at which $E_2^+ = E^+(k_{z2})$ increases via
the decay of AWs at $|k_z|= k_{z3}$ is 
\begin{equation}
R_{3\rightarrow 2}\sim \frac{k_{z2} E^+_2k_{z3} E^-_3}{\beta^{1/2}
  v_{\rm A}},
\label{eq:R32} 
\end{equation} 
where $E^-_3 = E^-(k_{z3})$,
while the rate at which $E_2^+ $ decreases via
the decay of AWs at $|k_z|= k_{z2}$ is 
\begin{equation}
R_{2\rightarrow 1}\sim \frac{k_{z2} E^+_2k_{z1} E^-_1}{\beta^{1/2}
  v_{\rm A}},
\label{eq:R21} 
\end{equation} 
where $E^-_1 = E^-(k_{z1})$. The time derivative of $E_2^+$ is
$R_{3\rightarrow 2} - R_{2 \rightarrow 1}$, or
\begin{equation}
\frac{\partial E^+_2}{\partial t} \sim \frac{k_{z2} E^+_2 (k_{z3}E^-_3 - k_{z1}E^-_1)}{\beta^{1/2}
  v_{\rm A}} .
\label{eq:diff} 
\end{equation} 
Equation~(\ref{eq:omegares3}) implies that $k_{z3} - k_{z1} \sim
 k_{z2} c_{\rm s}/v_{\rm A} \sim  \beta^{1/2} k_{z2}$. A Taylor
expansion of $k_{z3} E^-_3$ and $k_{z1} E^-_1$ about $k_{z2}$ in
Equation~(\ref{eq:diff}) thus allows this equation to be rewritten as
\begin{equation}
\frac{\partial E^+_2}{\partial t} \sim \frac{k_{z2}^2 E^+_2}{v_{\rm A}}
\frac{\partial}{\partial k_z} \left(k_z E^-)\right|_{k_z = k_{z2}},
\label{eq:approx} 
\end{equation} 
which is the same as Equation~(\ref{eq:dEpmdt}) to within a factor
of order unity. 

To be clear, no independent derivation is being presented for
Equations~(\ref{eq:R32}) and (\ref{eq:R21}). The foregoing discussion
merely points out that Equations~(\ref{eq:R32}) and (\ref{eq:R21}) are
equivalent (up to a factor of order unity) to
Equation~(\ref{eq:dEpmdt}), which is derived on the basis of weak
turbulence theory. It is worth pointing out, however, that several
features of Equations~(\ref{eq:R32}) and (\ref{eq:R21}) make sense on
a qualitative level. If either $E^+=0$ or $E^-=0$, then
$R_{3\rightarrow 2} =R_{2\rightarrow 1} = 0$, because the parametric
instability is a stimulated decay, which ceases if initially all the
AWs travel in the same direction. For fixed $E^+$ and~$E^-$,
$R_{3\rightarrow 2}$ and $R_{2\rightarrow 1}$ vanish as
$v_{\rm A} \rightarrow \infty$, since the fractional nonlinearities
vanish in this limit.  Also, $R_{3\rightarrow 2}$ and
$R_{2 \rightarrow 1}$ are proportional to~$\beta^{-1/2}$ (when
$S^\pm_k$ is negligibly small, as assumed) because the
parametric-decay contribution to $\partial A_k^\pm /\partial t$ is an
integral (over $\bm{p}$ and~$\bm{q}$) of third-order correlation
functions such as
$\langle \delta \bm{v}_k \cdot \delta \bm{B}_q \delta n_{p}\rangle$,
where $\delta \bm{v}_k$ and $\delta \bm{B}_q$ are the velocity and
magnetic-field fluctuations associated with AWs at wave
vectors~$\bm{k}$ and~$\bm{q}$, and $\delta n_p$ is the density
fluctuation associated with the slow waves at wave vector~$\bm{p}$
that are driven by the beating of the AWs at wave vectors~$\bm{k}$
and~$\bm{q}$.  For fixed AW amplitudes and fixed~$B_0$ and~$v_{\rm A}$, this driven density fluctuation is proportional to
$\beta^{-1/2}$, because as $\beta$ decreases the thermal pressure is
less able to resist the compression along~$\bm{B}_0$ resulting from
the Lorentz force that arises from the beating of the AWs.

\section{Linear Growth of the Parametric Instability}
\label{sec:linear} 

In the linear regime of the parametric instability, the spectrum of AWs
propagating in one direction, say~$E^+$, is taken to be fixed, and
$E^- \ll E^+$.  Equation~(\ref{eq:dEpmdt}) then implies that
$E^-$ increases exponentially in time with growth rate 
\begin{equation}
\gamma^- =
\frac{\pi k_z^2 }{v_{\rm A}} \frac{\partial }{\partial k_z} \left( k_z
  E^+\right).
\label{eq:gammalin} 
\end{equation} 
Equation~(\ref{eq:gammalin}) is equivalent to Equation~(18) of
\cite{cohen74} given the different normalizations of the AW power
spectra in the two equations. For example, Equation~(\ref{eq:Epmnorm})
implies that
$\int_{0}^\infty E^+ dk_z 
= (1/2)\int_{-\infty}^\infty E^+ dk_z = \langle |\delta \bm{B}|^2\rangle
/4\pi \rho$
when $E^- \ll E^+$, which can be compared with the un-numbered but
displayed equation under Equation~(9) of \cite{cohen74}.  As in the
present paper, \cite{cohen74} assumed that slow waves are strongly
damped and that the AWs satisfy the random-phase approximation. The
present paper builds upon the results of \cite{cohen74} by
investigating the coupled nonlinear evolution 
of $E^+$ and $E^-$. Also, whereas \cite{cohen74} took the wave vectors
to be parallel or anti-parallel to~$\bm{B}_0$, the derivation of
Equation~(\ref{eq:dEpmdt}) in the present
paper allows for obliquely propagating waves.

\section{Conservation of Wave Quanta and Inverse Cascade}
\label{sec:inverse} 

To simplify the presentation, it is assumed that
\begin{equation}
k_z > 0.
\label{eq:kzpos} 
\end{equation} 
No generality is lost, because $E^\pm$
is an even function of~$k_z$, and thus it is sufficient to solve for
the spectra at positive~$k_z$ values.
Equation~(\ref{eq:dEpmdt}) can be rewritten as the two equations
\begin{equation}
\frac{\partial N}{\partial t} +  \frac{\partial \Gamma}{\partial k_z}= 0
\label{eq:dNdt} 
\end{equation} 
and
\begin{equation} 
\frac{\partial \Gamma}{\partial t} = \pi \hbar k_z^2 \Gamma
\frac{\partial}{\partial k_z} \left(k_z^2 N \right),
\label{eq:dGammadt} 
\end{equation} 
where 
\begin{equation}
N = \frac{E^+ + E^-}{\hbar k_z v_{\rm A}}
\label{eq:defN} 
\end{equation} 
is the number of wave quanta per unit $k_z$ per unit mass and 
\begin{equation}
\Gamma = - \frac{\pi k_z^2 E^+ E^-}{\hbar v_{\rm A}^2}
\label{eq:defGamma}  
\end{equation} 
is the flux of wave quanta in $k_z$-space. Equation~(\ref{eq:dNdt}) implies that the
number of wave quanta per unit mass,
\begin{equation}
N_{\rm tot} = \int_{-\infty}^\infty N dk_z,
\label{eq:Ntot} 
\end{equation} 
is conserved. The fact that $\Gamma$ is negative indicates that there
is an inverse cascade of wave quanta from large~$k_z$ to
small~$k_z$ \citep[c.f.][]{terasawa86}.
The wavenumber drift velocity of the wave quanta,
\begin{equation}
\left \langle\frac{d k_z}{dt} \right \rangle \equiv \frac{\Gamma}{N} =
- \frac{\pi k_z^3}{v_{\rm A}}\left(\frac{1}{E^+} + \frac{1}{E^-}\right)^{-1},
\label{eq:dkzdt} 
\end{equation} 
is determined primarily by the smaller of~$E^+$ and~$E^-$.

\section{Exact Solutions to the Wave Kinetic Equations}
\label{sec:exact} 

In this section, several exact solutions to Equation~(\ref{eq:dEpmdt})
are presented under the assumption that $k_z > 0$. The spectra at
negative $k_z$ follow from the relation
$E^\pm(-k_z,t) = E^\pm(k_z,t)$. 

\subsection{Decaying, Balanced Turbulence}
\label{sec:Decaying} 

One family of exact solutions to Equation~(\ref{eq:dEpmdt}) follows
from setting
\begin{equation}
E^\pm(k_z,t) = f^\pm(k_z, t) H\big(k_z - b(t)\big)
\label{eq:Efb} 
\end{equation} 
in Equation~(\ref{eq:dEpmdt}),
where
\begin{equation}
H(x) = \left\{\begin{array}{ll}
0 & \mbox{ if $x<0$} \\
1 & \mbox{ if $x\geq 0$}
\end{array}
\right.
\label{eq:heaviside} 
\end{equation} 
is the Heaviside function. When Equation~(\ref{eq:Efb}) is substituted
into Equation~(\ref{eq:dEpmdt}), each side of
Equation~(\ref{eq:dEpmdt}) becomes the sum of terms proportional
to~$\delta(k_z-b)$ and terms that contain no delta function.  By
separately equating the two groups of terms, one can show that
Equation~(\ref{eq:Efb}) is a solution to Equation~(\ref{eq:dEpmdt}) if
\begin{equation}
\frac{\partial }{\partial t}f^\pm(k_z,t) = \frac{\pi}{v_{\rm A}}k_z^2 f^\pm(k_z,t)
\frac{\partial}{\partial k_z}\left[k_z f^\mp(k_z,t)\right]
\label{eq:feq} 
\end{equation} 
and
\begin{equation}
\frac{1}{b^3} \frac{db}{dt} = -\frac{\pi f^+(b,t)}{2v_{\rm A}} = 
-\frac{\pi f^-(b,t)}{2v_{\rm A}} .
\label{eq:dbdt0} 
\end{equation} 
Equation~(\ref{eq:dbdt0}) makes use of the relation $[H(x)]^2 = H(x)$
and its derivative, $2 H(x) \delta (x) = \delta(x)$.  In
Appendix~\ref{ap:BL} it is shown that Equation~(\ref{eq:dbdt0}) can be
recovered by adding a small amount of nonlinear diffusion to
Equation~(\ref{eq:dEpmdt}) and replacing the discontinuous jump in
the spectrum at $k_z = b(t)$ with a boundary
layer. Equation~(\ref{eq:dbdt0}) implies that, for solutions of the
form given in Equation~(\ref{eq:Efb}), the mean-square amplitudes of
forward and backward-propagating AWs must be equal just above the
break wavenumber~$b$. An exact solution to Equations~(\ref{eq:feq}) and (\ref{eq:dbdt0})
corresponding to decaying turbulence is 
\begin{equation}
f^+(k_z,t) = f^-(k_z,t) =  \frac{a(t)}{k_z^2},
\label{eq:fpdecay} 
\end{equation} 
\begin{equation}
a(t) = a_0 \left(1 + \frac{\pi a_0 t}{v_{\rm A}}\right)^{-1},
\label{eq:adecay} 
\end{equation}
and
\begin{equation}
b(t) = b_0 \left(1 + \frac{\pi a_0 t}{v_{\rm A}}\right)^{-1/2},
\label{eq:bdecay} 
\end{equation}
where $a_0$ and $b_0$ are the values of $a$ and~$b$ at $t=0$.

This solution can be further truncated at large~$k_z$ by setting
\begin{equation}
E^+(k_z,t) = E^-(k_z,t) = \frac{a(t) H(k_z - b(t)) H(q(t) - k_z)}{k_z^2}
\label{eq:trunc2} 
\end{equation} 
with 
\begin{equation}
q(t) = q_0\left(1+ \frac{\pi a_0 t}{v_{\rm A}}\right)^{-1/2},
\label{eq:defq} 
\end{equation} 
where $q_0$ is the value of~$q$ at $t=0$, which is taken to
exceed~$b_0$. Equations~(\ref{eq:fpdecay})
through (\ref{eq:defq}) can be recovered numerically by solving
Equation~(\ref{eq:dEpmdt}) for freely decaying AWs.
Whether the spectra satisfy
Equations (\ref{eq:Efb}) and (\ref{eq:fpdecay}) through~(\ref{eq:bdecay}) 
or, alternatively,  Equations~(\ref{eq:adecay}) through~(\ref{eq:defq}), the number of wave
quanta $N_{\rm tot}$ defined in Equation~(\ref{eq:Ntot}) is
finite and independent
of time.

\subsection{Forced, Balanced Turbulence}
\label{sec:Forced} 

An exact solution to Equations~(\ref{eq:feq}) and (\ref{eq:dbdt0})
corresponding to forced turbulence is
\begin{equation}
f^+(k_z,t) = f^-(k_z,t) = \frac{c}{k_z}
\label{eq:fsteady} 
\end{equation} 
and
\begin{equation}
b(t) = \left( \frac{\pi c t}{2v_{\rm A}} + \frac{1}{b_0}\right)^{-1},
\label{eq:bsteady} 
\end{equation} 
where $c$ is a constant and $b_0$ is the value of~$b$ at~$t=0$.
In this solution, the number of wave quanta~$N_{\rm tot}$ is not
constant, because there is a nonzero influx of wave quanta from infinity.
A version of this solution can be realized in a numerical solution of
Equation~(\ref{eq:dEpmdt}) by holding
$E^\pm$ fixed at some wavenumber $k_{\rm f}$, which mimics the effects
of energy input from external forcing. In this case, the numerical solution at $k_z < k_{\rm f}$
is described by Equations~(\ref{eq:Efb}), (\ref{eq:fsteady}),
and~(\ref{eq:bsteady}), with $b(t) < k_{\rm f}$.

The solution in Equations~(\ref{eq:fsteady}) and
(\ref{eq:bsteady}) can be truncated at large~$k_z$ in a manner
analogous to Equation~(\ref{eq:trunc2}), but with
$q = [(\pi c t/2v_{\rm A}) + (1/q_0)]^{-1}$, where $q_0$ is the value
of~$q$ at $t=0$. In this solution, $N_{\rm tot}$ is independent of
time. Numerical solutions of
Equation~(\ref{eq:dEpmdt}) show, however, that this solution is
unstable. If the spectra initially satisfy
$E^\pm = (c/k_z) H(k_z - b)H(q-k_z)$, then they evolve towards the
solution described by Equations~(\ref{eq:fpdecay}) through
(\ref{eq:defq}).

\subsection{Exact Solutions Extending over All~$k_z$}
\label{sec:Allkz} 

In addition to the truncated solutions described in
Sections~\ref{sec:Decaying} and~\ref{sec:Forced},
Equation~(\ref{eq:dEpmdt})
possesses several exact solutions  that
extend over all~$k_z$. These solutions are unphysical, because
they correspond to infinite AW energy and neglect dissipation (which
becomes important at sufficiently large~$k_z$) and finite system size
(which becomes important at sufficiently small~$k_z$). However, they
illustrate several features of the nonlinear evolution of the
parametric instability, which are summarized at the end of this
section.

The simplest solution to
Equation~(\ref{eq:dEpmdt}) spanning all~$k_z$ is
\begin{equation}
E^\pm(k_z,t) = \frac{c^\pm}{k_z},
\label{eq:Epsteady} 
\end{equation} 
where $c^\pm$ is a constant. It follows from 
Equation~(\ref{eq:defGamma}) that Equation~(\ref{eq:Epsteady}) 
corresponds to a constant flux of
AW quanta to smaller~$k_z$. 
In contrast to the truncated $E^\pm \propto k_z^{-1}$
forced-turbulence solution in
Section~\ref{sec:Forced}, $E^+$ and $E^-$ need not be equal in
Equation~(\ref{eq:Epsteady}).

A second, non-truncated, exact solution to Equation~(\ref{eq:dEpmdt})
is given by 
\begin{equation} 
E^\pm(k_z,t)  =  \frac{a^\pm(t)}{k_z^2}
\label{eq:Eunbdec}  
\end{equation}
and
\begin{equation} 
a^\pm(t) =  \frac{a_0^\pm (a_0^\pm - a_0^\mp)}{a_0^\pm - a_0^\mp e^{-\pi(a_0^\pm -
    a_0^\mp)t/v_{\rm A}}} ,
\label{eq:apm1} 
\end{equation} 
where $a_0^+$ and $a_0^-$ are the initial values of $a^+$ and~$a^-$.
In this solution,
\begin{equation}
a^+(t) - a^-(t) = a_0^+ - a_0^-.
\label{eq:adiff} 
\end{equation} 
If $a^+_0 > a^-_0$, then $E^-$ decays faster than $E^+$, and, after a
long time has passed, $E^-$ decays to zero while $a^+$ decays to the
value $a^+_0 - a^-_0$.  Conversely, if $a^-_0 > a_0^+$, then $E^+$
decays faster than~$E^-$, and the turbulence decays to a state in
which $E^+= 0$.  In the limit that
$a_0^+ \rightarrow a_0^-$,
\begin{equation}
a^\pm(t) \rightarrow a_0\left(1 + \frac{\pi a_0 t}{v_{\rm A}}\right)^{-1},
\label{eq:abal} 
\end{equation} 
where $a_0 = a^+_0 = a^-_0$. Equations~(\ref{eq:Eunbdec}) and
(\ref{eq:abal}) are a non-truncated version of the decaying-turbulence
solution presented in Section~\ref{sec:Decaying}.

Equations~(\ref{eq:Epsteady}) and (\ref{eq:Eunbdec}) 
can be combined into a more general class of
solution, 
\begin{equation}
E^\pm(k_z,t) = a^\pm(t) \left(\frac{1}{k_z^2} + \frac{d^\pm}{k_z}\right),
\label{eq:comb1} 
\end{equation} 
where $d^+$ and $d^-$ are constants and $a^\pm(t)$ is given by
Equation~(\ref{eq:apm1}). Another type of solution combining
$k_z^{-1}$ and $k_z^{-2}$ scalings is
\begin{eqnarray} 
E^+(k_z,t) & = & \frac{c_0 e^{-\pi c_2 t/v_{\rm A}}}{k_z}, \\
E^-(k_z,t) & = & \frac{c_1}{k_z} + \frac{c_2 }{k_z^2},
\label{eq:12} 
\end{eqnarray} 
where $c_0$, $c_1$, and~$c_2$ are constants.

The exact solutions presented in this section illustrate three properties of the nonlinear
evolution of the parametric instability at low~$\beta$ when slow waves
are strongly damped. First, when $E^\pm \propto k_z^{-1}$,
$\partial E^\mp/\partial t$~vanishes.  Second, if
$E^\pm \propto k_z^{-2}$, then $(\partial/\partial t) \ln E^\mp$ is
negative and independent of~$k_z$, and $E^\mp(k_z, t)$ can be written
as the product of a function of $k_z$ and a (decreasing) function of
time.  (More general principles describing the evolution of~$E^\pm$
are summarized in Figure~\ref{fig:slope_evolution} and
Equation~(\ref{eq:slope_evol}).)  Third, the parametric instability
does not necessarily saturate with $E^+ = E^-$. For example, in
Equations~(\ref{eq:Eunbdec}) and (\ref{eq:apm1}), when
$a_0^+ \neq a_0^-$, the AWs decay to a maximally aligned state
reminiscent of the final state of decaying cross-helical
incompressible MHD turbulence \citep{dobrowolny80}.

\section{Nonlinear Evolution of the Parametric Instability When
  Most of the AWs Initially Propagate in the Same Direction}
\label{sec:SW} 

This section describes a numerical solution to Equation~(\ref{eq:dEpmdt})
in which, initially, 
\begin{equation}
E^+ \gg E^-.
\label{eq:EpggEm} 
\end{equation} 
As in Section~\ref{sec:exact}, $k_z$ is taken to be positive, and the
spectra at negative $k_z$ can be inferred from the fact that
$E^\pm(-k_z) = E^\pm(k_z)$.  The spectra are advanced forward in time
using a second-order Runge-Kutta algorithm on a logarithmic wavenumber
grid consisting of 2000 grid points.  To prevent the growth of
numerical instabilities, a nonlinear diffusion term
\begin{equation}
D^\pm = \nu E^\mp k_z^2 \frac{\partial^2}{\partial k_z^2}E^\pm
\label{eq:defDpm} 
\end{equation} 
is added to the right-hand side of Equation~(\ref{eq:dEpmdt}), where
$\nu$ is a constant. The value of
$\nu$ is chosen as small as possible subject to the constraint
that the diffusion term suppress instabilities at the grid scale.

To represent the solution in a way that can be readily compared with
spacecraft measurements of solar-wind turbulence,
the wavenumber spectra are converted into frequency spectra,
\begin{equation}
e^\pm(f,t) = \frac{2\pi E^\pm(k_z, t)}{U},
\label{eq:defef} 
\end{equation} 
where $U$ is the solar-wind
velocity, and
\begin{equation}
f = \frac{k_z U}{2\pi}
\label{eq:deff} 
\end{equation} 
is the frequency in the spacecraft frame that, according to Taylor's~(\citeyear{taylor38})
hypothesis, corresponds to wavenumber~$k_z$ 
when the background magnetic field is aligned with the nearly radial
solar-wind velocity.  The Alfv\'en speed is
taken to be the approximate average of the observed values of~$v_{\rm A}$
in three fast-solar-wind streams at $r=0.3 \mbox{ AU}$ (see
Table~1 of \cite{marsch82a}  and Table 1a of \cite{marsch90}),
\begin{equation}
v_{\rm A} = 150 \mbox{ km/s}.
\label{eq:va} 
\end{equation} 
In order to compare directly with Figure 2-2c of \cite{tumarsch95},
the  solar-wind velocity is taken to be
\begin{equation}
U = 733 \mbox{ km/s}.
\label{eq:U} 
\end{equation} 

The power spectra are initialized to the values
\begin{equation}
e^+(f, t=0) = \frac{\sigma^+(f/f_0)^{-0.5}}{1 + (f/f_0)^{1.5}}
\label{eq:Epinit} 
\end{equation} 
and
\begin{equation}
e^-(f, t=0) = \sigma^-,
\label{eq:Eminit} 
\end{equation} 
where $\sigma^+$, $\sigma^-$, and~$f_0$ are constants. 
The values of~$f_0$ and the corresponding wavenumber~$k_{z0}$ are chosen so that 
\begin{equation}
f_0 = \frac{k_{z0} U}{2\pi} = 10^{-2} \mbox{ Hz},
\label{eq:valkz0} 
\end{equation} 
consistent with the arguments of \cite{vanballegooijen16} about the
dominant frequency of AW launching by the Sun. The minimum and maximum
wavenumbers of the numerical domain are chosen so that $k_{z\rm max} =
10^3 k_{z0} = 10^7 k_{z\rm min}$.  The motivation for the scaling
$e^+(f, t=0) \propto f^{-0.5}$ at small~$f$ is the similar scaling
observed by \cite{tumarsch95} in the aforementioned fast-solar-wind
stream at $10^{-5} \mbox{ Hz} < f < 10^{-4} \mbox{ Hz}$. The numerical
results shown below suggest that the parametric instability has little
effect on $e^+$ at these frequencies at $r=0.3 \mbox{ AU}$. The
observed $f^{-0.5}$ scaling in this frequency range is thus presumably
inherited directly from the spectrum of AWs launched by the Sun. Like
the scaling $e^+ \propto f^{-0.5}$, the
value of~$\sigma^+$ is chosen to match the observed spectrum of
outward-propagating AWs at 0.3~AU at small~$f$.  The reason for the $f^{-2}$
scaling in $e^+$ at large~$f$ is that a (parallel) $k_z^{-2}$ spectrum
is observed in the solar wind~\citep{horbury08, podesta09c,forman11}
and predicted by the theory of critically balanced MHD turbulence
\citep[see, e.g.,][]{goldreich95,mallet15}.  The value of $\sigma^-$
is set equal to a minuscule value ($10^{-12} \sigma^+$), so that the
only source of dynamically important, inward-propagating AWs is the
parametric decay of outward-propagating AWs.

\begin{figure}
\centerline{
\includegraphics[width=6cm]{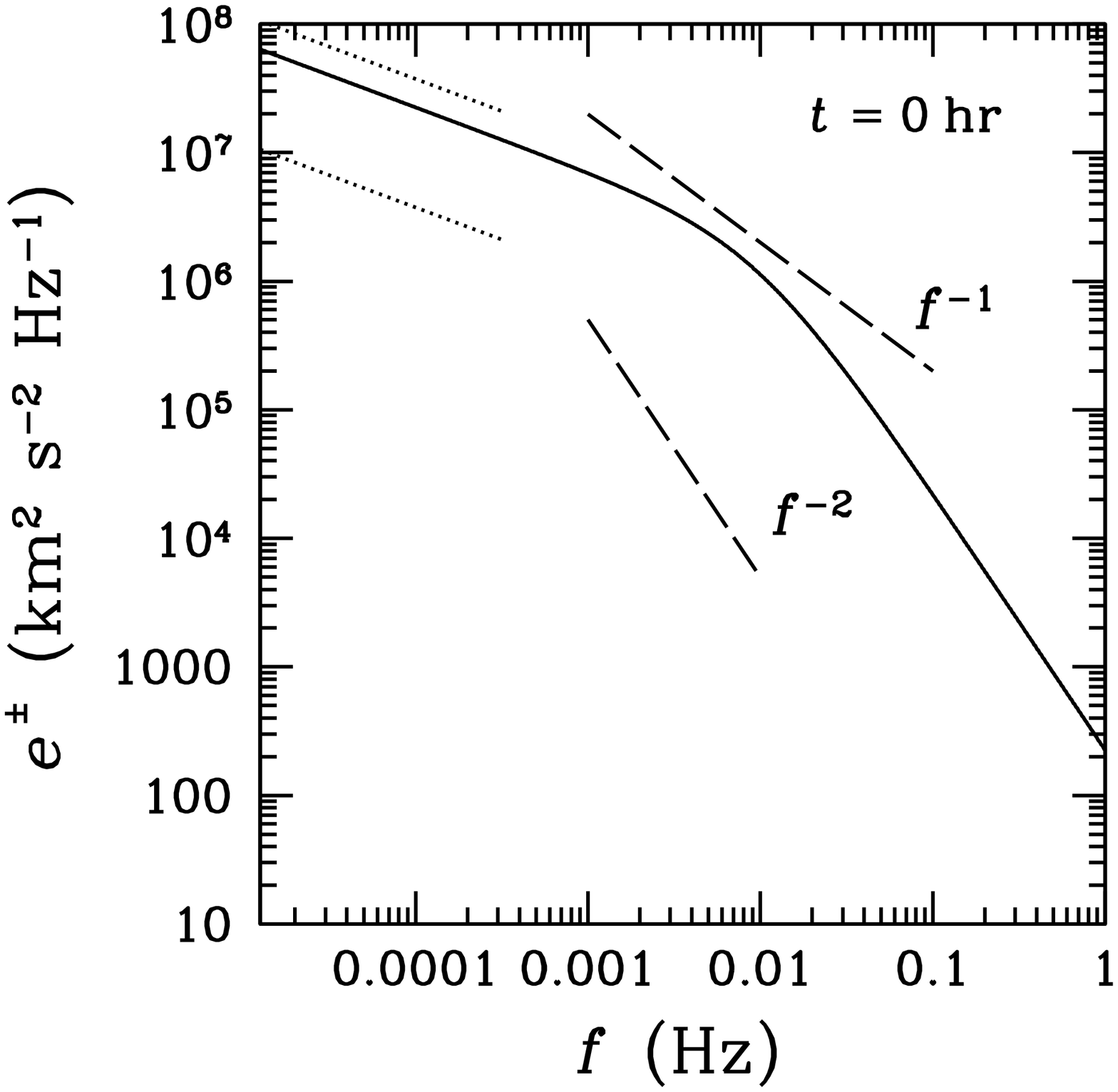}
\includegraphics[width=6cm]{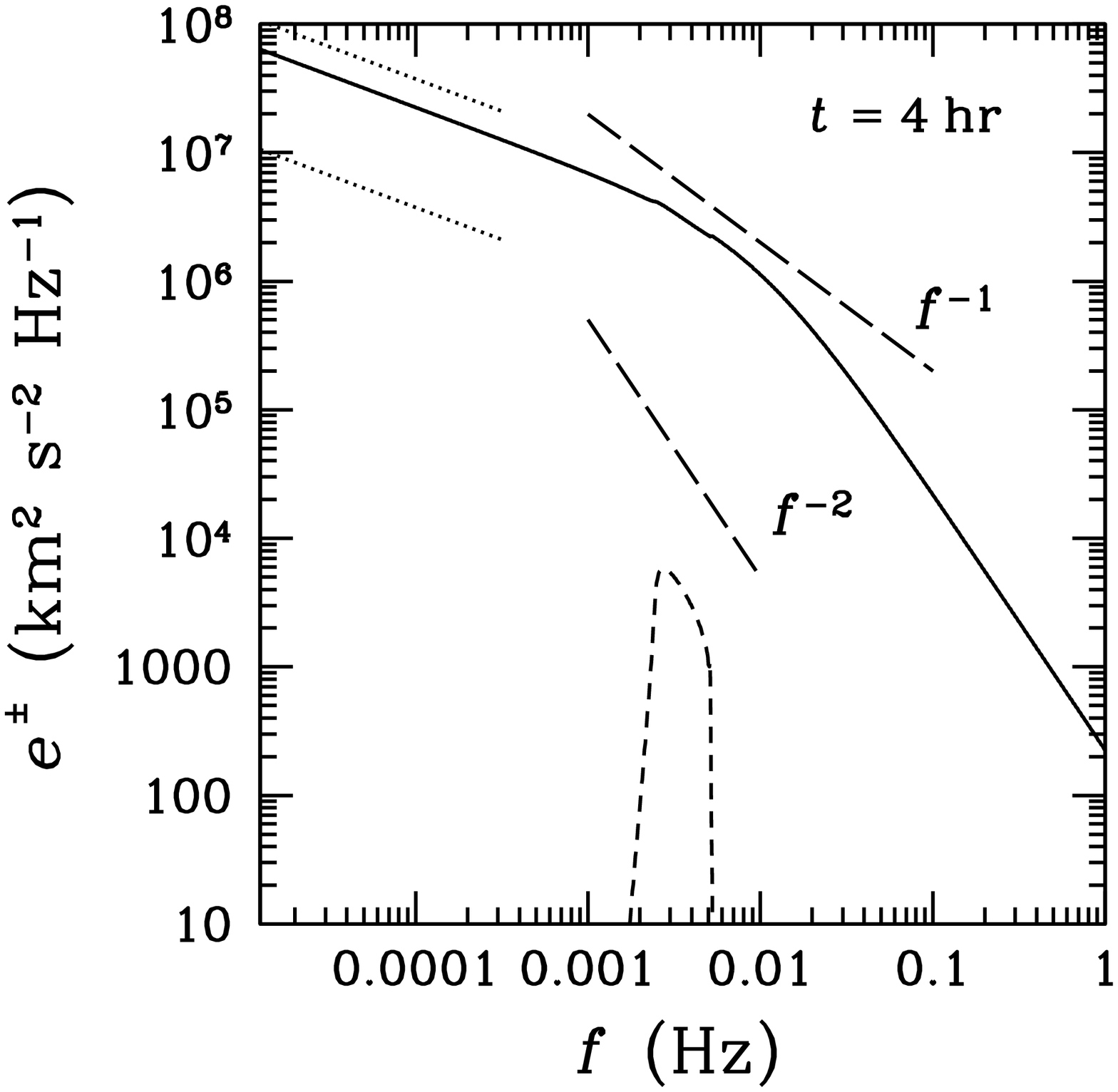}
}
\centerline{
\includegraphics[width=6cm]{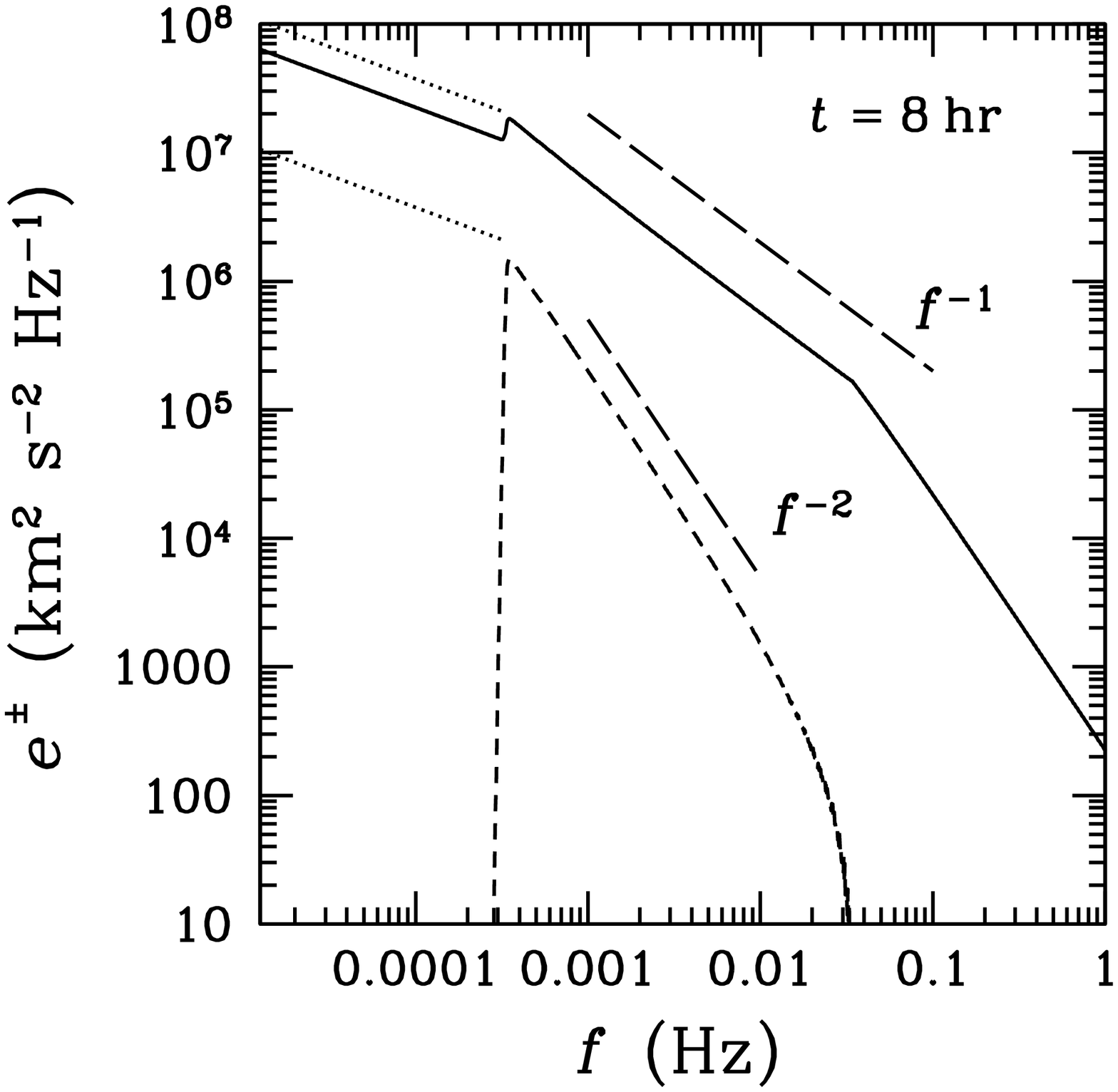}
\includegraphics[width=6cm]{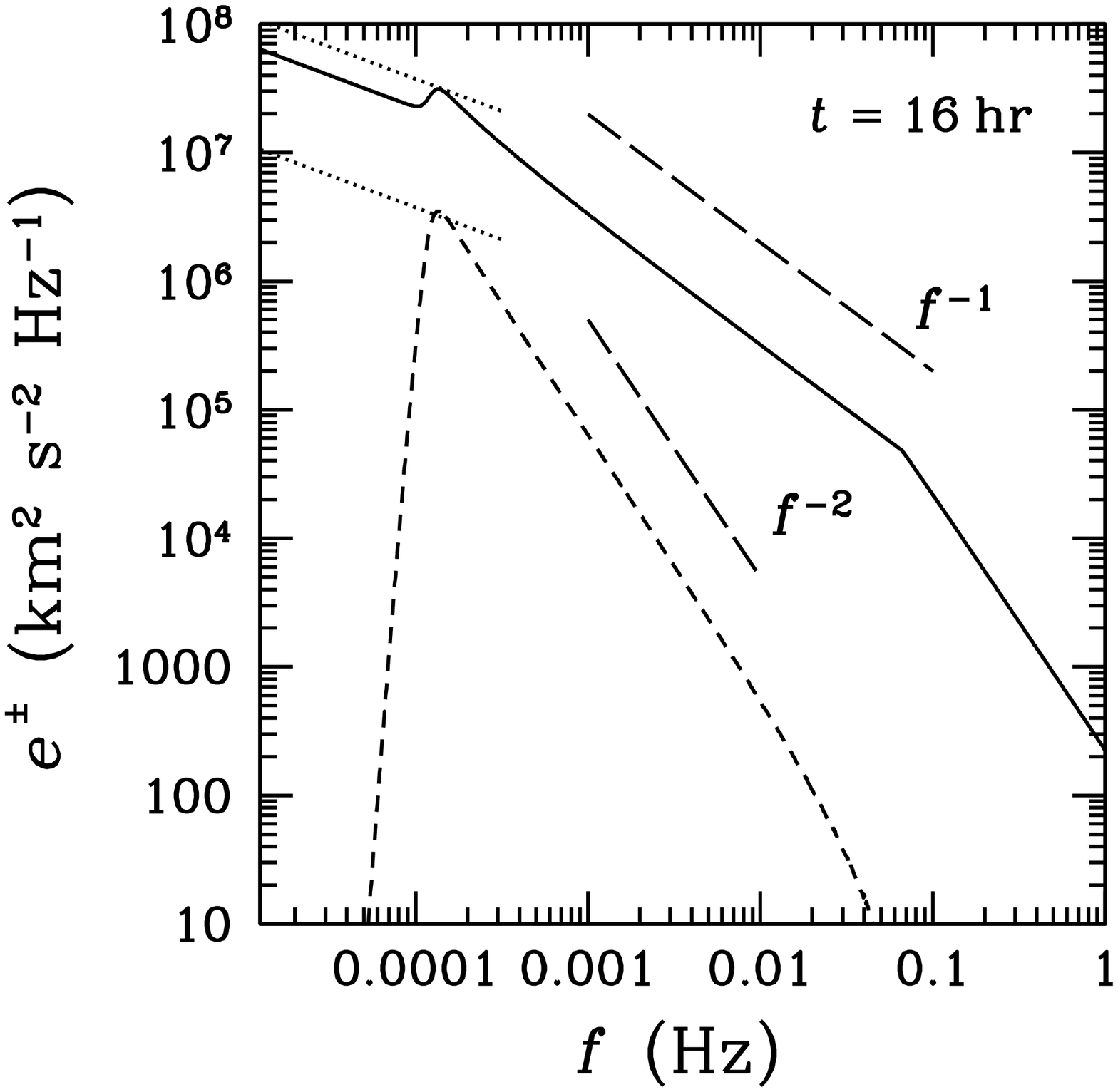}
}
\caption{Solid lines show the AW power spectra in a numerical solution of
  Equation~(\ref{eq:dEpmdt}) with plasma parameters and turbulence
  parameters chosen to model conditions in the fast solar wind at a
  heliocentric distance of 0.3~AU.  The wavenumber spectra
  $E^\pm(k_z)$ appearing in Equation~(\ref{eq:dEpmdt}) have been
  converted, using Equations~(\ref{eq:defef}) and (\ref{eq:deff}),
  into the frequency spectra~$e^\pm(f)$.  The dotted lines in the
  upper left corners of each plot show the evolutionary tracks of the
  values of~$e^+$ and $e^-$ at the low-frequency end of the frequency range
  in which $e^+ \propto f^{-1}$  in the approximate analytic solution
  to Equation~(\ref{eq:dEpmdt}) presented in Appendix~\ref{ap:approx}.
 \label{fig:num_wide}}
\end{figure}

Figure~\ref{fig:num_wide} summarizes the results of the calculation.
Between $t=0$ and $t= 4 \mbox{ hr}$, $e^+$ changes little while $e^-$
grows rapidly between roughly 2 and 5~mHz, where the growth
rate~$\gamma^-$ given in Equation~(\ref{eq:gammalin}) peaks.  Between
$t=4 \mbox{ hr}$ and $t=8 \mbox{ hr}$, $e^+$ develops a broad
$\sim 1/f$ scaling between $f= 3 \times 10^{-4} \mbox{ Hz}$ and
$f = 3 \times 10^{-2} \mbox{ Hz}$, which shuts off the growth of $e^-$
at these frequencies. At the same time, $e^-$ acquires an
$\sim f^{-2}$ scaling over much of this same frequency range. Between
$t= 8 \mbox{ hr}$ and $t=16 \mbox{ hr}$, the low-frequency limit of
the $1/f$ range of~$e^+$ decreases to $\sim 10^{-4} \mbox{ Hz}$, and the
high-frequency limit of the $1/f$ range of~$e^+$ increases to
$\sim 0.1 \mbox{ Hz}$.

The dotted lines in the upper left corner of each panel in
Figure~\ref{fig:num_wide} show the tracks followed by
the values of $e^+$ and~$e^-$ at the low-frequency end of the
frequency range in which $e^+ \propto f^{-1}$ in the approximate analytic
solution to Equation~(\ref{eq:dEpmdt}) that is described in
Appendix~\ref{ap:approx}. In this solution, $E^+$ and $E^-$ are
expanded in negative powers of~$k_z$ at wavenumbers exceeding a
time-dependent break wavenumber~$b(t)$.
Below this wavenumber, $E^-=0$ and $E^+ = \eta k_z^p$, where
$\eta$ and~$p$ are constants, and $-1 < p < 1$.  At $k_z > b$, the dominant term in
the expansion of~$E^+$~($E^-$) scales like $k_z^{-1}$~($k_z^{-2}$),
and the ratios of~$E^+(b_+)$ to~$\eta b_+^p$ and $E^+(b_+)$ to~$E^-(b_+)$ are fixed
functions of~$p$, where $b_+$ is a wavenumber infinitesimally larger than~$b$.
For~$p=-0.5$, $E^+(b_+)/\eta b_+^p = 5/3$ and
$E^+(b_+)/E^-(b_+) = 10$, in approximate agreement with the numerical
results (see also the right panel of Figure~\ref{fig:num_narrow}).

\subsection{Heuristic Explanation of the $e^+ \propto f^{-1}$ and
  $e^- \propto f^{-2}$ Scalings}
\label{sec:model} 

\begin{figure}
\centerline{
\includegraphics[width=10cm]{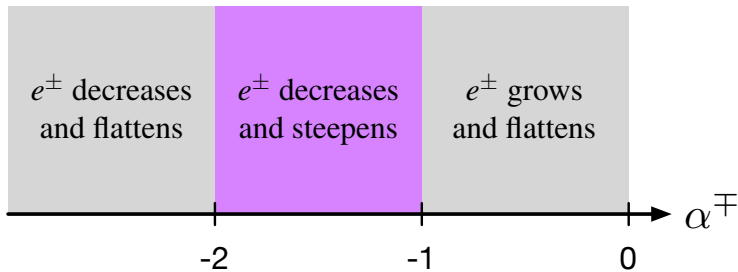}
}
\caption{In this figure, it is assumed that the frequency spectra are
  initially power laws of the form
$e^\pm \propto f^{\alpha^\pm}$, and that $\alpha^+$ and $\alpha^-$ are
both negative. According to Equation~(\ref{eq:slope_evol}),  parametric decay
  alters both the amplitude and slope of $e^\pm$ in the
  manner shown. For example, if $e^- \propto f^{-1.5}$, then $E^+
  \propto k_z^{-1.5}$, and Equation~(\ref{eq:slope_evol}) implies
  that $E^+$ decreases at a rate that increases with~$k_z$. This in
  turn implies that $e^+$
  decreases at a rate that increases with~$f$, so that $e^+$ steepens.
  \label{fig:slope_evolution}}
\end{figure}

In order to understand the time evolution illustrated in
Figure~\ref{fig:num_wide}, it is 
instructive to first consider the case in which
\begin{equation}
E^\pm = c^\pm k_z^{\alpha^\pm}
\label{eq:alphapm} 
\end{equation} 
within some interval $(k_{z1}, k_{z2})$, where $c^\pm$ and $\alpha^\pm$ are constants.
Equation~(\ref{eq:dEpmdt}) implies that, within this interval,
\begin{equation}
\frac{\partial}{\partial t} \ln E^\pm = \frac{\pi c^\mp}{v_{\rm A}}\left(
  1 + \alpha^\mp\right) k_z^{\alpha^\mp + 2}.
\label{eq:slope_evol} 
\end{equation} 
If $\alpha^\mp > -1$, then
$\ln E^\pm$ grows at a rate that increases with~$k_z$, causing
$E^\pm$ to increase and ``harden,'' in the sense that the best-fit
value of $\alpha^\pm$ within the interval
$(k_{z1}, k_{z2})$ increases. 
 If $\alpha^\mp = -1$, then
$E^\pm$ does not change.  If $-2 < \alpha^\mp < -1$, then $\ln E^\pm$
decreases at a rate that increases with~$k_z$, which causes
the best-fit value of~$\alpha^\pm$ within the interval $(k_{z1},
k_{z2})$ to decrease.  If $\alpha^\mp = -2$,
then $\ln E^\pm$ decreases at the same rate at all~$k_z$, and 
$\alpha^\pm$ remains unchanged. Finally, if
$\alpha^\mp < 2$, then $E^\pm$ decreases at a rate that decreases
with~$k_z$, which causes the best-fit value of~$\alpha^\pm$ in
the interval $(k_{z1}, k_{z2})$ to increase.
These rules are summarized in
Figure~\ref{fig:slope_evolution} and apply to
$e^\pm \propto f^{\alpha^\pm} $ as well as $E^\pm \propto k_z^{\alpha^\pm}$.

Returning to Figure~\ref{fig:num_wide}, in the early stages of the
numerical calculation, $e^-$ grows most rapidly at those frequencies
at which $\gamma^-$ in Equation~(\ref{eq:gammalin}) is largest ---
namely, the high-$f$ end of the $f^{-0.5}$ range of $e^+$. By the time
$e^-$ reaches a sufficient amplitude that $e^+$ and $e^-$ evolve on
the same timescale, $e^-$ develops a peaked frequency profile
extending from some frequency $f = f_{\rm low}$ to some larger
frequency $f = f_{\rm high}$, as illustrated in the upper-right panel
of Figure~\ref{fig:num_wide}. Near $f_{\rm low}$, $de^-/df > 0$ (i.e.,
$\alpha^- > 0$), which causes $e^+$ to grow.\footnote{Although the
  caption of Figure~\ref{fig:slope_evolution} excludes
  positive~$\alpha^\mp$ to allow use of the words ``flattens'' and
  ``steepens'' in the figure, Equation~(\ref{eq:slope_evol}) applies
  for positive~$\alpha^\mp$.}  Near $f_{\rm high}$, $\alpha^- < - 1$,
which causes $e^+$ to decrease. Thus, $e^+$ steepens across the
interval $(f_{\rm low}, f_{\rm high})$ until it attains a $1/f$
scaling, at which point $e^-$ stops growing between $f_{\rm low}$ and
$f_{\rm high}$. However, at frequencies just below $f_{\rm low}$,
$e^+$ and $e^-$ both continue to grow, causing $f_{\rm low}$ to
decrease. At the same time, $e^+$ continues to decrease at larger $f$
where $\alpha^- < -1$. Together, the growth of $e^+$ just below
$f_{\rm low}$ and the damping of $e^+$ at larger~$f$ cause the
$f^{-1}$ range of $e^+$ to broaden in both directions, i.e., towards
both smaller and larger frequencies.

The unique scaling of $e^-$ consistent with an $e^+$ spectrum
$\propto f^{-1}$ that is a decreasing function of time is
$e^- \propto f^{-2}$. Moreover, the scalings $e^+ \sim f^{-1}$ and
$e^- \sim f^{-2}$ are, in a sense, stable, as can be inferred from
Figure~\ref{fig:slope_evolution}.  For example, if $\alpha^-$
increases from $-2$ to a slightly larger value, then $e^+$ decreases
at a rate that increases with~$f$, causing $\alpha^+$ to decrease to a
value slightly below~$-1$. This causes $e^-$ to decrease at a rate
that increases with~$f$, thereby causing $\alpha^-$ to decrease back
towards~$-2$. A similar ``spectral restoring force'' arises for any
other small perturbation to the values $\alpha^+ = -1$ and
$\alpha^- = -2$.

\begin{figure}
\centerline{
\includegraphics[trim = 4cm 0cm 0cm 0cm, width=4.5cm]{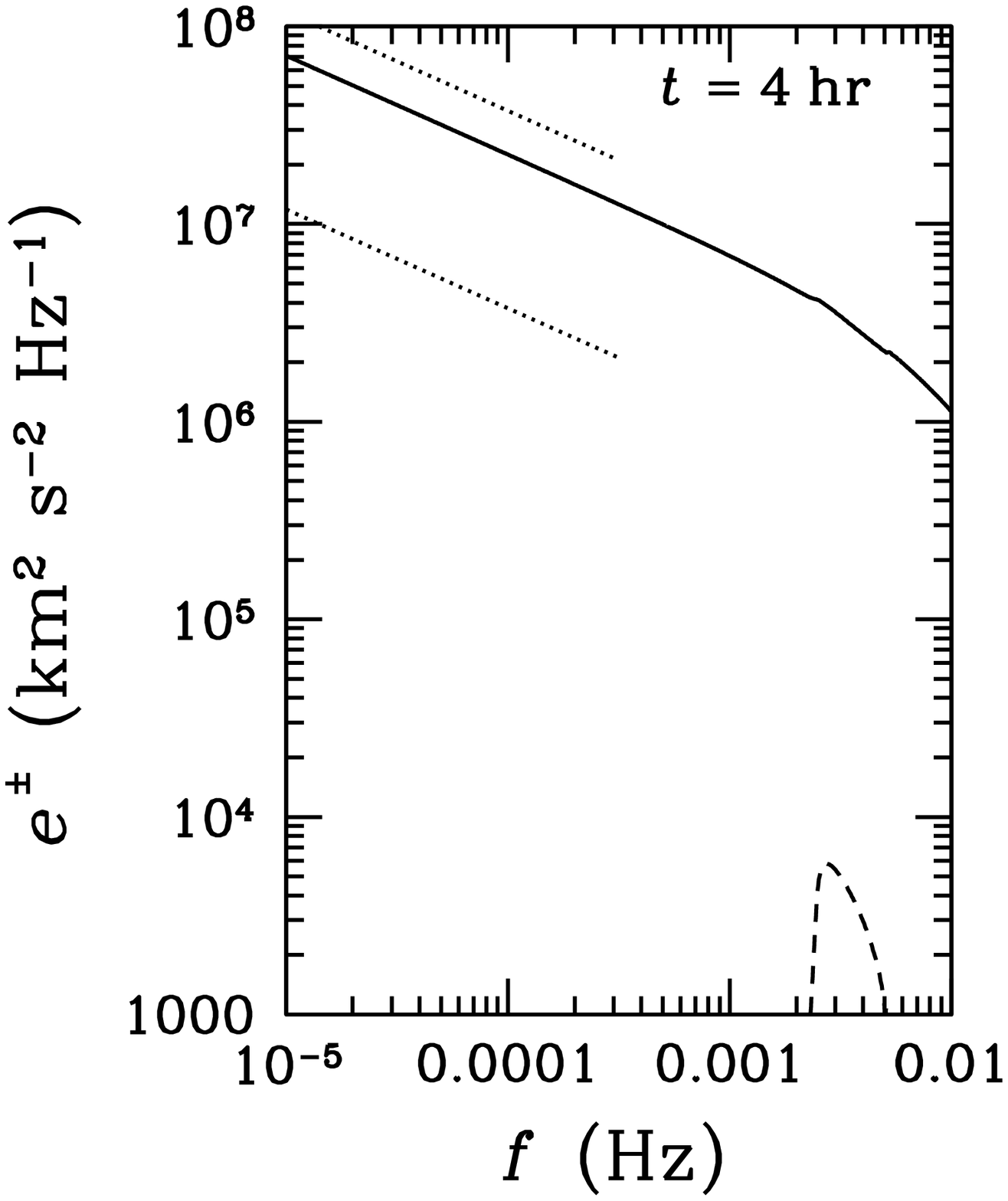}
\includegraphics[trim = 4cm 0cm 0cm 0cm, width=4.5cm]{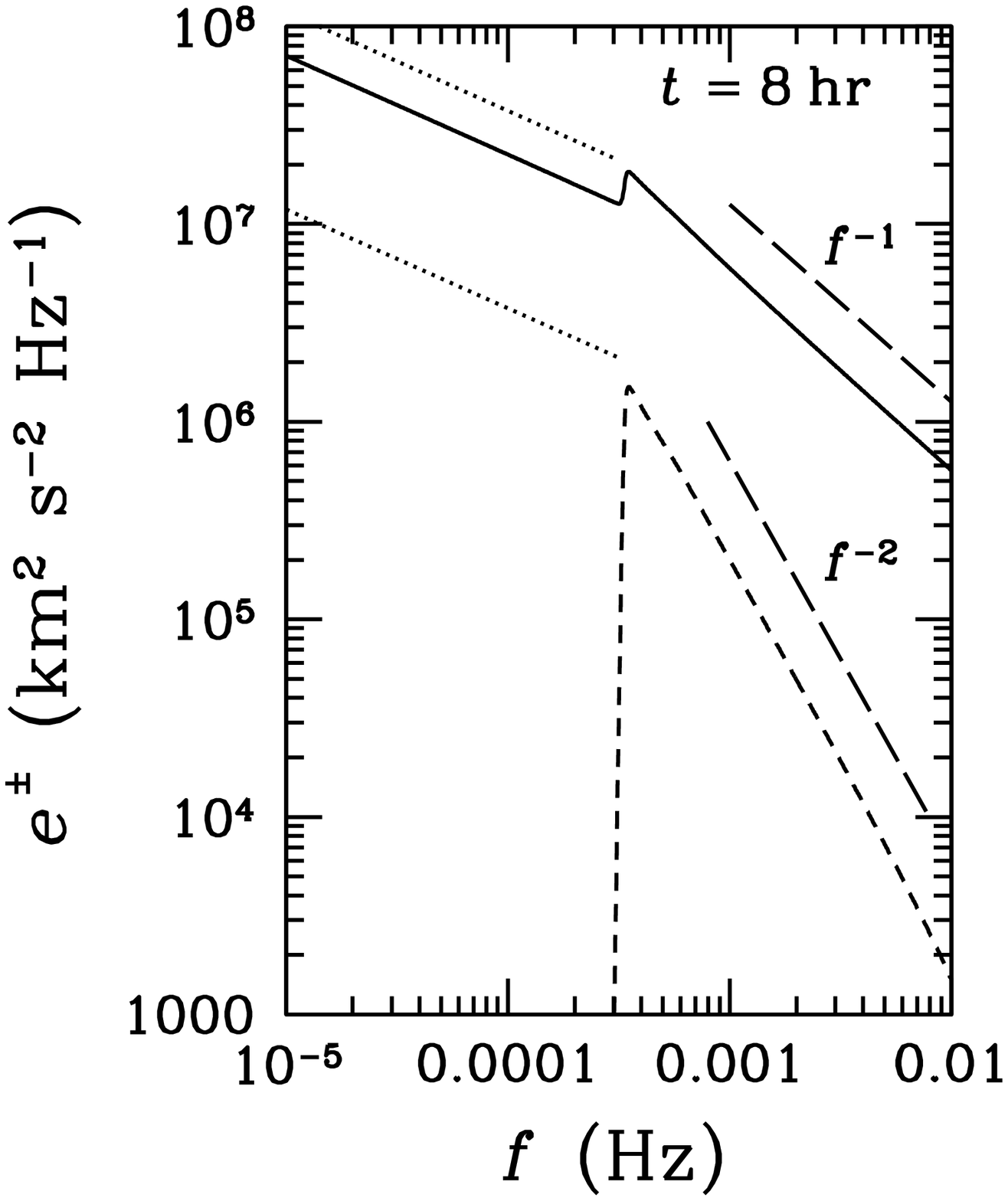}
\includegraphics[trim = 4cm 0cm 0cm 0cm, width=4.5cm]{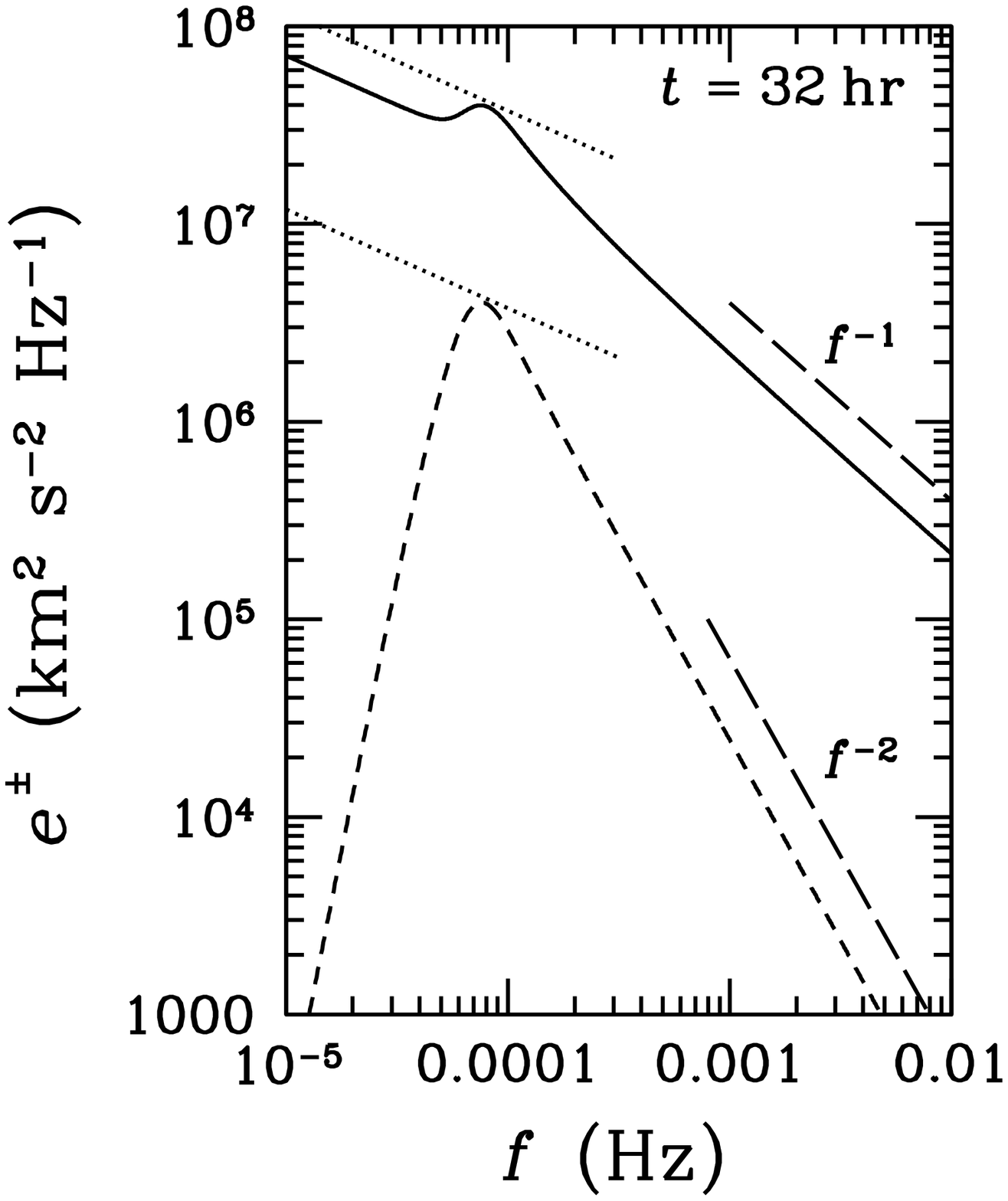}
}
\caption{The left panel and middle panel of this figure reproduce the
  $t=4 \mbox{ hr}$ and $t= 8 \mbox{ hr}$ panels of
  Figure~\ref{fig:num_wide} but with the axis ranges used in
  Figure~2-2c of \cite{tumarsch95}. The right panel is from a later
  time ($t=32 \mbox{ hr}$) in the same numerical solution.
 \label{fig:num_narrow}}
\end{figure}

It is worth emphasizing, in this context, that the analytic solution
presented in Appendix~\ref{ap:approx} is approximate rather than
exact. As the spectral break frequency decreases past some fixed
frequency $f_{3}$, the values of $e^+$ and $e^-$ at $f_{3}$ suddenly
jump, but they do not jump to the precise values needed to extend the
$e^+ \sim f^{-1}$ and $e^- \sim f^{-2}$ scalings to
smaller~$f$. Instead, the spectra need further ``correcting'' after
the break frequency has swept past in order to maintain the scalings
$e^+ \sim f^{-1}$ and $e^- \sim f^{-2}$ in an approximate way.  Also,
the decrease in $e^-$ that occurs after $t=4 \mbox{ hr}$ is a
consequence of the sub-dominant $f^{-2}$ component of~$e^+$. This
component of~$e^+$ becomes increasingly prominent near the break
frequency as time progresses, leading to the pronounced curvature in
the plot of $e^+$ near $f=10^{-4} \mbox{ Hz}$ in the right panel of
Figure~\ref{fig:num_narrow}.

\subsection{Comparison with {\em Helios} Measurements}
\label{sec:comp} 

In the (average) plasma rest frame,
the equations of incompressible MHD can be written in
the  form
\begin{equation}
  \frac{\partial \bm{z}^\pm}{\partial t} + ( \bm{z}^\mp\pm \bm{v}_{\rm A}) \cdot \nabla
\bm{z}^\pm = - \nabla \Pi,
\label{eq:elsasser} 
\end{equation} 
where $\bm{z}^\pm = \delta \bm{v} \mp \delta \bm{B}/\sqrt{4\pi \rho}$
are the Elsasser variables, $\delta \bm{v}$ and $\delta \bm{B}$ are
the velocity and magnetic-field fluctuations, $\rho$ is the mass
density, $\bm{v}_{\rm A} = \bm{B}_0/\sqrt{4\pi \rho}$ is the Alfv\'en
velocity, and $\Pi$ is the total pressure divided by~$\rho$
\citep{elsasser50}. Although the solar wind is compressible,
Equation~(\ref{eq:elsasser}) provides a reasonable approximation for
the non-compressive, AW-like component of solar-wind turbulence.
As Equation~(\ref{eq:elsasser}) shows, the advection
velocity of a $\bm{z}^\pm$ fluctuation is $\bm{z}^\mp \pm \bm{v}_{\rm
  A} $.  This implies, as shown by \cite{maron01}, that
$\bm{z}^\pm$ fluctuations propagate
along magnetic field lines perturbed by $\bm{z}^\mp$.
As a consequence, in the solar wind, when the rms
magnetic-field fluctuation $\delta B_{\rm in}$ associated with
inward-propagating AWs~($\bm{z}^-$) is much smaller than the
background magnetic field~$B_0$, the outward-propagating
AWs~($\bm{z}^+$) propagate to a good approximation along the direction
of~$\bm{B}_0$. This is true even if the rms magnetic-field
fluctuation~$\delta B_{\rm out}$ associated with $\bm{z}^+$ is
comparable to~$B_0$. In the fast solar wind at $r< 0.3 \mbox{ AU}$,
the (fractional) cross helicity is high (i.e., $E^+ \gg E^-$), and
$\delta B_{\rm in}$ is indeed small compared to~$B_0$
\citep{bavassano00,cranmer05}. Moreover, the background magnetic field
at $r=0.3 \mbox{ AU}$ is nearly in the radial direction, because the
Parker-spiral magnetic field begins to deviate appreciably from the
radial direction only at larger~$r$ in the fast wind
\citep{verscharen15}. Hence, in high-cross-helicity fast-wind streams
at $r=0.3 \mbox{ AU}$, the function $e^+$ defined by
Equations~(\ref{eq:defef}) and (\ref{eq:deff}) corresponds to a good
approximation to the frequency spectrum of outward-propagating AWs
observed by a spacecraft in the solar wind. It is not clear, however,
how well $e^-$ corresponds to the observed spectrum of
inward-propagating AWs, because the inward-propagating AWs follow
field lines perturbed by the outward-propagating AWs, which can be
inclined relative to the radial direction by a substantial angle.

Figure~\ref{fig:num_narrow} reproduces the $t=4 \mbox{ hr}$ and
$t=8 \mbox{ hr}$ panels of Figure~\ref{fig:num_wide}, but with the
same axis ranges as those in Figure 2-2c of \cite{tumarsch95} to
facilitate comparison. Figure~\ref{fig:num_narrow} also includes a
third panel that shows the spectra at $t=32 \mbox{ hr}$.  The $e^+$
spectrum in the $t=8 \mbox{ hr}$ panel of Figure~\ref{fig:num_wide}
shares a number of properties with the $e^+$ spectrum in Figure 2-2c
of \cite{tumarsch95}, in addition to the $f^{-0.5}$ scaling at
small~$f$ that was built in to the numerical calculation as an initial
condition.  In particular, $e^+ \gg e^-$ at all frequencies,
$e^+ \sim f^{-1}$ at $f\gtrsim 3 \times 10^{-4} \mbox{ Hz}$, and there
is a bump in the $e^+$ spectrum at the transition between the
$f^{-0.5}$ and $f^{-1}$ scaling ranges of~$e^+$.

Although this comparison is suggestive, it is not entirely clear how
to map time in the numerical calculation to heliocentric distance in
the solar wind, because the plasma parameters in the numerical
calculation are independent of position and time, whereas they depend
strongly upon heliocentric distance in the solar wind. For example,
the turbulence is weaker (in the sense of smaller $\delta v_0/v_{\rm
  A}$) the closer one gets to the Sun. (See also the discussion
following Equation~(\ref{eq:kzLperp}).) Also, the choice of initial
conditions in the numerical calculation artificially prolongs the
linear stage of evolution, since in the solar wind there are sources
of inward-propagating waves other than parametric instability, such as
non-WKB reflection \citep{heinemann80,velli93}. Nevertheless, as a
baseline for comparison, the travel time of an outward-propagating AW
from the photosphere to $0.3 \mbox{ AU}$ in the fast-solar-wind model
developed by \cite{chandran09c} is approximately 12~hr.

\section{Discussion of Approximations and Relevance to the Solar Wind}
\label{sec:applicability} 

This section critically assesses the assumptions underlying
the results in Sections~\ref{sec:WKE} through~\ref{sec:SW} and the
degree to which these assumptions apply to the fast solar wind between
$r=10 R_{\odot}$ (the approximate perihelion of the {\em Parker Solar
  Probe}) and $r=0.3 \mbox{ AU}$.  

\subsection{The Weak Turbulence Approximation}
\label{sec:weak} 

A central assumption of the analysis is the
weak-turbulence criterion in Equation~(\ref{eq:wto}). Since $E^+$ and $E^-$
differ in the solar wind, Equation~(\ref{eq:wto}) is really two
conditions,
\begin{equation}
\omega_{\rm nl}^\pm \ll |k_z| v_{\rm A},
\label{eq:wto1} 
\end{equation} 
where $\omega_{\rm nl}^+$ ($\omega_{\rm nl}^-$) is the inverse of the
timescale on which nonlinear
interactions modify outward-propagating (inward-propagating) AWs.
The contribution to $\omega_{\rm nl}^\pm$ from the parametric instability is
\begin{equation}
\omega_{\rm nl, PI}^\pm \sim  \frac{1}{E^\pm} \left|\frac{\partial
  E^\pm}{\partial t}\right| \sim
\frac{k_z^2 E^\mp}{v_{\rm A}} .
\label{eq:omegaPI} 
\end{equation} 
The contribution to $\omega_{\rm nl}^\pm$ from one other type of
nonlinear interaction is estimated in Section~\ref{sec:NL}.
The estimate of $\partial E^\pm/\partial t$ in
Equation~(\ref{eq:omegaPI}) follows 
from Equation~(\ref{eq:dEpmdt}) and setting
$E^\mp \sim |k_z|^{\alpha^\mp}$ with $\alpha^\mp$ not very close
to~$-1$. A rough upper limit on $\omega_{\rm nl, PI}^\pm$ results from
replacing
$k_z E^\mp$ in Equation~(\ref{eq:omegaPI}) with $(\delta v^\mp)^2$, where
$(\delta v^+)^2$  is the mean-square velocity fluctuation associated
with outward-propagating AWs, and
$(\delta v^-)^2$  is the mean-square velocity fluctuation associated
with inward-propagating AWs. This leads to a rough upper limit on
$\omega_{\rm nl, PI}^\pm$ because
$(\delta v^\pm)^2$ includes contributions 
from all wavenumbers and is much larger than
the value of $k_z E^\pm$ at some~$k_z$.  Equation~(\ref{eq:wto1}),
with $\omega_{\rm nl} \sim \omega_{\rm nl, PI}$, 
is thus satisfied provided
\begin{equation}
(\delta v^\mp)^2\ll v_{\rm A}^2.
\label{eq:wto2} 
\end{equation} 

\cite{bavassano00} analyzed {\em Helios} measurements of fluctuations
in the fast solar wind at $r= 0.4 \mbox{ AU}$ and found that
$(\delta v^-)^2 \ll (\delta v^+)^2 \simeq (60 \mbox{ km/s})^2$. As
mentioned above, the typical value of $v_{\rm A}$ in the fast solar
wind at $r= 0.3 \mbox{ AU}$ is $\sim 150 \mbox{ km/s}$
\citep{marsch82a,marsch90}. Near $r=0.3 \mbox{ AU}$, $B_0 \sim 1/r^2$,
$\rho \sim 1/r^2$, and $v_{\rm A} \sim 1/r$, and so the typical value
of $v_{\rm A}$ in fast-solar-wind streams at $r=0.4 \mbox{ AU}$ is
$\sim 112.5 \mbox{ km/s}$.  These measurements indicate that
\begin{equation}
(\delta v^-)^2 \ll (\delta v^+)^2 \simeq 0.28 v_{\rm A}^2
\label{eq:wto3} 
\end{equation} 
in the fast solar wind at $r=0.4 \mbox{ AU}$. Since $\delta
v^\pm/v_{\rm A}$ decreases as $r$ decreases below 0.4~AU
\citep{cranmer05,chandran09c}, the condition $\omega_{\rm nl, PI}^+
\ll |k_z| v_{\rm A}$ is well
satisfied at $r< 0.4 \mbox{ AU}$, and the condition
$\omega_{\rm nl, PI}^- \ll |k_z| v_{\rm A}$ is at least marginally satisfied
at $r< 0.4 \mbox{ AU}$.

It is worth noting that weak turbulence theory fails when applied to resonant
interactions between three AWs, because such interactions occur only
when one of the AWs has zero frequency, violating the weak-turbulence
ordering~\citep{schekochihin12,meyrand15}.  In contrast, the AW/slow-wave
interactions in parametric decay
do not involve a zero-frequency mode. Weak turbulence
theory is thus in principle a better approximation for the nonlinear
evolution of the parametric instability than for incompressible MHD turbulence.

\subsection{The Low-$\beta$ Assumption}
\label{sec:lowbeta} 

The assumption that $\beta \ll 1$ is not 
satisfied at $r \gtrsim 0.3 \mbox{ AU}\simeq 65 R_{\odot}$, where $\beta$ is~typically
$\sim 1$,
but is reasonable at $r \lesssim 20 R_{\odot}$ \citep{chandran11}. It
is possible that the $\beta \ll 1$ theory presented here applies at
least at a qualitative level provided $\beta$ is simply~$\lesssim 1$, and
indeed this possibility motivates the comparison of the present model with {\em
  Helios} observations. However, further work is
needed to investigate how the results of this paper are modified
as $\beta$ increases to values $\sim 1$.

\subsection{Neglect of Other Types of Nonlinear Interactions}
\label{sec:NL} 

Another approximation in 
Sections~\ref{sec:WKE} through~\ref{sec:SW}   is the neglect of
all nonlinear interactions
besides parametric decay. One of the neglected 
interactions is the shearing of inward-propagating AWs by
outward-propagating AWs, which makes a contribution to $\omega^-_{\rm
  nl}$ that depends on the perpendicular length scale of the AWs. At the
perpendicular outer scale~$L_\perp$ (the overall correlation length of the AWs
measured perpendicular to~$\bm{B}_0$), the contribution to
$\omega^-_{\rm nl}$ from shearing is approximately
\begin{equation}
\omega_{\rm nl, \perp}^- \sim \frac{ \chi \delta v^+}{L_\perp},
\label{eq:omegaperp1} 
\end{equation}
where 
\begin{equation}
\chi = \frac{\delta v^+}{k_z L_\perp v_{\rm A}}
\label{eq:defchi} 
\end{equation} 
is the critical-balance parameter 
\citep{goldreich95,ng96,lithwick07}. Equation~(\ref{eq:omegaperp1}) does not apply
when $\chi$ is much larger than~1, but direct numerical simulations
suggest that $\chi \lesssim 1$ at $r \gtrsim 10 R_{\odot}$ for the
bulk of the AW energy (J. Perez, private communication). Thus, at $r \gtrsim
10 R_{\odot}$,
\begin{equation}
\frac{\omega_{\rm nl, PI}^-}{\omega_{\rm nl, \perp}^-} \simeq (k_z L_\perp)^2.
\label{eq:nlcomp} 
\end{equation} 

As AWs propagate away from the Sun, they
follow magnetic field lines, which leads
to the approximate scaling $L_\perp \propto B_0^{-1/2}$. In the WKB limit, the AW
frequency in the Sun's frame 
$k_z (U+v_{\rm A})$ is independent of~$r$. The scaling
$k_z \sim 1/(U+ v_{\rm A})$ thus serves as a rough approximation for
outward-propagating AWs in the turbulent solar wind. At the coronal base (just above the
transition region), where $k_z$ and $L_\perp$ have the values $k_{z\rm b}$
and $L_{\perp \rm b}$, the value of $ k_{z\rm b} L_{\perp \rm b}$ for
the energetically dominant AWs launched by the Sun can
be estimated (in essence from the critical-balance condition) as  $\delta v_{\rm
  b}^+/v_{\rm Ab}$, where $\delta v_{\rm b}^+$ and $v_{\rm A b}$ are the
values of $\delta v^+$ and $v_{\rm A}$ at the coronal base \citep{goldreich95,vanballegooijen16}.
Together, these scalings lead to the estimate 
\begin{equation}
k_z L_\perp \simeq \sqrt{\frac{B_{0\rm b}}{B_0}} \left(\frac{ U_{\rm b} +
    v_{\rm A b}}{U + v_{\rm A}}\right)\frac{\delta
  v^+_{\rm b}}{v_{\rm A b}},
\label{eq:est1} 
\end{equation} 
where $B_{0\rm b}$ and $U_{\rm b}$ are the values of the background
magnetic field and solar-wind outflow velocity at the coronal
base. Between $r=10 R_{\odot}$ and $r= 60 R_{\odot}$, $B_{0\rm b}/B_0
\simeq f_{\rm max} (r/R_{\odot})^2$, where $f_{\rm max}$ is the super-radial
expansion factor \citep{kopp76}. In the fast solar wind within this range of radii, $U+v_{\rm
  A} \simeq 700 - 800 \mbox{ km/s}$, which is comparable to $U_{\rm b}
+ v_{\rm A b} \simeq v_{\rm Ab} \simeq 10^3 \mbox{
  km/s}$. Equation~(\ref{eq:est1}) is thus approximately equivalent to
\begin{equation}
k_z L_\perp \simeq  \sqrt{f_{\rm max}}\,
\left(\frac{r}{R_{\odot}}\right) \left(\frac{\delta v_{\rm b}^+}{v_{\rm A
    b}}\right)
\label{eq:est2} 
\end{equation} 
for the energetically dominant fluctuations launched by the Sun.
If we set $f_{\rm max} = 9$, $\delta v_{\rm b} = 30 \mbox{ km/s}$, and
$v_{\rm A b} = 900 \mbox{ km/s}$, then Equation~(\ref{eq:est2})
becomes
\begin{equation}
k_z L_\perp \sim \frac{r}{10 R_{\odot}}
\label{eq:kzLperp} 
\end{equation} 
for the energetically dominant AWs launched by the
Sun. Equations~(\ref{eq:nlcomp}) and (\ref{eq:kzLperp}) suggest that
it is reasonable to neglect the shearing of inward-propagating AWs by
outward-propagating AWs at $r \gtrsim 10 R_{\odot}$. On the other
hand, at smaller radii, shearing could suppress the growth of
inward-propagating AWs that would otherwise result from the parametric
instability. Also, the requirement that $k_z L_\perp > 1$ in order
for~$\omega_{\rm nl, PI}^-$ to exceed~$\omega_{\rm nl, \perp}^-$ could
prevent the $f^{-1}$ range from spreading to frequencies below some
minimum ($r$-dependent) value.

The other nonlinearities in the weak-turbulence wave kinetic equations
that are neglected in this paper include interactions involving fast
magnetosonic waves, the turbulent mixing of slow waves by AWs, phase
mixing of AWs by slow waves, and the shearing of outward-propagating
AWs by inward-propagating AWs \citep{chandran08b}. In-situ
measurements indicate that fast waves account for only a small
fraction of the energy in compressive fluctuations at 1~AU
\citep{yao11,howes12,klein12}. Also, fast waves propagating away from
the Sun undergo almost complete reflection before they can escape into
the corona \citep{hollweg78}. These findings suggest that nonlinear
interactions involving fast waves have little effect upon the
conclusions of this paper. The turbulent mixing of slow waves by AWs
acts as an additional slow-wave damping mechanism and is thus unlikely
to change the conclusions of this paper, which already assume strong
slow-wave damping.  Phase mixing of AWs by slow waves transports AW
energy to larger~$k_\perp$ at a rate that increases
with~$|k_z|$~\citep{chandran08b}.  Although the fractional density
fluctuations between $r = 10 R_{\odot}$ and $r= 0.3 \mbox{ AU}$ are
fairly small \citep[see, e.g.,][]{tumarsch95,hollweg10}, phase mixing
could affect the parallel AW power spectra, and further work is needed
to investigate this possibility.  The shearing of outward-propagating
AWs by inward-propagating AWs is enhanced by non-WKB reflection, which
makes this shearing more coherent in time \citep{velli89}. The
resulting nonlinear timescale for outward-propagating AWs is roughly
$r/(U+v_{\rm A})$ \citep{chandran09c}, where $U$ is the solar-wind
outflow velocity. This timescale is comparable to the AW propagation
time from the Sun to heliocentric distance~$r$, and hence to the
parametric-decay timescale at the small-$f$ end of the $1/f$ range
of~$e^+$. How this shearing modifies~$E^+(k_z)$, however, is not
clear. For example, shearing by inward-propagating AWs may transport
outward-propagating-AW energy to
larger~$k_\perp = \sqrt{k_x^2 + k_y^2}$ at a rate that is independent
of~$|k_z|$, in which case this shearing would reduce $E^+(k_z)$ by
approximately the same factor at all~$k_z$, leaving the functional
form of~$E^+(k_z)$ unchanged. 

\subsection{Neglect of Spatial Inhomogeneity}
\label{sec:expansion} 

In this paper, it is assumed that the background plasma is uniform and stationary. In the
solar wind, however, as an AW propagates from the
low corona to 0.3~AU, the properties of the ambient plasma seen by the
AW change dramatically, with $\beta$ increasing from~$\sim 10^{-2}$
to~$\sim 1$ and $\delta v_{\rm rms}/v_{\rm A}$ increasing from
$\sim 0.02$ to~$\sim 0.5$ \citep[][]{bavassano00,cranmer05,chandran11}. Further work is
needed to determine how this spatial inhomogeneity affects the
nonlinear evolution of the
parametric instability.

\subsection{Approximate Treatment of Slow-Wave Damping}
\label{sec:damping} 

A key assumption in Sections~\ref{sec:WKE} through~\ref{sec:SW} is
that slow waves are strongly damped, and this damping is
implemented by neglecting terms in the wave
kinetic equations that are proportional to the slow-wave power
spectrum~$S^\pm_k$. There are two sources of error in this
approach. First, damping could modify the polarization properties of
slow waves, thereby altering the wave kinetic equations. Second, even
if $S^\pm_k$ is much smaller than the AW power spectrum~$A^\pm_k$, the
neglected parametric-decay terms in the wave kinetic equations for AWs
that are proportional to~$S^\pm_k$ could still be important, because
they contain a factor of $\beta^{-1}$, which is absent in the
terms that are retained. This factor arises from the fact that the
fractional density fluctuation of a slow wave is $\sim \beta^{-1/2}$
times larger than the fractional magnetic-field fluctuation of an AW
with equal energy.  These neglected terms act to equalize the 3D AW
power spectra $A^+$ and~$A^-$, and hence to equalize $E^+$
and~$E^-$. If these neglected terms were in fact important, they could
invalidate the solutions presented in Section~\ref{sec:SW}, in
which~$E^+ \gg E^-$. However, in situ observations indicate that
$E^+ \gg E^-$ in the fast solar wind at $r= 0.3 \mbox{ AU}$
\citep{marsch90,tumarsch95}, which suggests that the neglect of these
terms is reasonable. Further work is
needed to investigate these issues more carefully.

\section{Conclusion}
\label{sec:conclusion} 

In this paper, weak turbulence theory is used to investigate the
nonlinear evolution of the parametric instability in low-$\beta$
plasmas. The analysis starts from the wave kinetic equations
describing the interactions between AWs and slow waves in weak
compressible MHD turbulence.  To account for the strong damping of
slow waves in collisionless plasmas, terms containing the slow-wave
energy density are dropped. The equations allow for all wave-vector
directions, but are integrated over the wave-vector components
perpendicular to the background magnetic field~$\bm{B}_0$ ($k_x$ and
$k_y$), which leads to equations for the 1D power spectra $E^+$
and~$E^-$ that depend only on the parallel wavenumber~$k_z$ and time.
During parametric decay in a low-$\beta$ plasma, an AW decays into a
slow wave propagating in the same direction and a counter-propagating
AW with a frequency slightly smaller than the frequency of the initial
AW.  The total number of AW quanta is conserved, and
the reduction in AW frequencies leads to an inverse cascade of AW
quanta towards smaller $\omega$ and~$k_z$.  The energy of each AW
quantum is $\hbar \omega$, and the decrease in $\omega$ during each
decay corresponds to a decrease in the AW energy, which is compensated
for by an increase in the slow-wave energy.  The subsequent damping
and dissipation of slow-wave energy results in plasma heating.

The main results of this paper concern the parametric decay of a
population of AWs propagating in one direction, say parallel
to~$\bm{B}_0$, when the counter-propagating AWs start out with much
smaller amplitudes. If the initial frequency spectrum~$e^+$ of the
parallel-propagating AWs has a peak frequency~$f_0$ (at which 
$f e^+$ is maximized) and an ``infrared'' scaling $f^p$ at smaller~$f$
with $-1 < p < 1$, then $e^+$ acquires a $1/f$ scaling throughout a
range of frequencies that spreads out in both directions
from~$f_0$. At the same time, the anti-parallel-propagating AWs
acquire a $1/f^2$ spectrum within this same frequency range. If the
plasma parameters and infrared $e^+$ spectrum are chosen to match
conditions in the fast solar wind at a heliocentric distance of
0.3~AU, and the AWs are allowed to evolve for a period of time that is
roughly two-thirds of the AW travel time from the Sun to 0.3~AU, the
resulting form of $e^+$ is similar to the form observed by the {\em
  Helios} spacecraft in the fast solar wind at 0.3~AU. Because the
background plasma parameters are time-independent in the analysis of
this paper but time-dependent in the plasma rest frame in the solar
wind, it is not clear how to map the time variable in the present
analysis to heliocentric distance.  Nevertheless, the similarity
between the spectra found in this paper and the spectra observed by
{\em Helios} suggests that parametric decay plays an
important role in shaping the AW spectra observed in the fast solar
wind at 0.3~AU, at least for wave periods $\lesssim 1 \mbox{ hr}$.

The frequency $f^\ast$ that dominates the AW energy is the maximum
of~$(fe^+ + fe^-)$. At the beginning of the numerical calculation
presented in Section~\ref{sec:SW}, $f^\ast$ is approximately
$f_0 = 0.01 \mbox{ Hz}$.  At $t = 8 \mbox{ hr}$ in this numerical
calculation, $f^\ast$ is the smallest frequency at
which~$e^+ \sim f^{-1}$ and $e^- \sim f^{-2}$, which
is~$\sim 3 \times 10^{-4} \mbox{ Hz}$.  This decrease in~$f^\ast$ is a
consequence of the aforementioned inverse cascade, which transports AW
quanta from the initial peak frequency to smaller frequencies. Inverse
cascade offers a way to reconcile the observed dominance of AWs at
hour-long timescales at 0.3~AU with arguments that the Sun launches
most of its AW power at significantly shorter wave periods
\citep{cranmer05,vanballegooijen16}.

Further work is needed to relax some of
the simplifying assumptions in this paper, including the
low-$\beta$ approximation, the assumption of spatial homogeneity, the
simplistic treatment of slow-wave damping, and the neglect of
nonlinear interactions other than parametric decay. Further work 
is also needed to evaluate the relative contributions of parametric
decay and other mechanisms to the generation of $1/f$ spectra in the
solar wind. For example, \cite{matthaeus86} argued that the $f^{-1}$
spectrum seen at $r=1 \mbox{ AU}$ at
$3\times 10^{-6} \mbox{ Hz} < f < 8\times 10^{-5} \mbox{ Hz}$ is a
consequence of forcing at the solar surface, and \cite{velli89} argued
that the shearing of
outward-propagating AWs by the inward-propagating AWs produced by
non-WKB reflection causes the outward-propagating AWs to acquire an
$f^{-1}$ spectrum.  

NASA's {\em Parker Solar Probe} (PSP) has a planned launch date in the
summer of 2018 and will reach heliocentric distances less
than~$10 R_{\odot} $. The FIELDS \citep{bale16} and SWEAP
\citep{kasper15} instrument suites on PSP will provide the first-ever
in-situ measurements of the magnetic-field, electric-field, velocity,
and density fluctuations in the solar wind at $r< 0.29 \mbox{ AU}$.
Although the issues mentioned in the preceding paragraph are sources
of uncertainty, the results of this paper lead to the following
predictions that will be tested by PSP. First, the $1/f$ range
of~$e^+$ in fast-solar-wind streams at $r < 0.3 \mbox{ AU}$ and
$f\gtrsim 3 \times 10^{-4} \mbox{ Hz}$ is produced in situ by 
parametric decay. As a
consequence, the $1/f$ range of~$e^+$ in fast-solar-wind streams will
be much more narrow at small~$r$ than at $r=0.3 \mbox{ AU}$. As AWs
propagate away from the Sun, the frequency range
$(f_{\rm min}, f_{\rm max})$ in which $e^+ \sim 1/f$ spreads out in
both directions from the near-Sun peak frequency (the maximum of
$f e^+$). Thus, $f_{\rm min}$ will be larger closer to the Sun,
and $f_{\rm max}$ will be~smaller.  Finally, during epochs in which
the local magnetic field is aligned with the relative velocity between
the plasma and the spacecraft (see the discussion in
Section~\ref{sec:comp}), the spectrum of $e^-$ will scale like~$1/f^2$
in the frequency interval $(f_{\rm min}, f_{\rm max})$.

I thank Phil Isenberg for discussions about the work of
\cite{cohen74} and the three anonymous reviewers for helpful comments
that led to improvements in the manuscript.  This work was supported
in part by NASA grants NNX15AI80, NNX16AG81G, and NNX17AI18G, NASA
grant NNN06AA01C to the Parker Solar Probe FIELDS Experiment, and NSF
grant PHY-1500041.

\appendix

\section{Boundary Layers in the Alfv\'en-Wave Power Spectra}
\label{ap:BL} 

In this appendix, 
a small nonlinear diffusion term is added to the wave kinetic equation,
so that Equation~(\ref{eq:dEpmdt}) becomes
\begin{equation}
\frac{\partial E^\pm}{\partial t} = \frac{\pi }{v_{\rm A}} k_z^2 E^\pm
\frac{\partial}{\partial k_z} \left( k_z E^\mp\right)+ \nu \overline{
  E^\mp} \frac{\partial ^2}{\partial k_z^2} E^\pm,
\label{eq:dEpmdt2} 
\end{equation} 
where $\nu$ is a constant that is taken to be very small,
\begin{equation}
\overline{ E^\mp}(k_z,t) = \frac{1}{2\Delta}\int_{k_z -
  \Delta}^{k_z+\Delta} E^\mp  dk_z,
\label{eq:avEmp} 
\end{equation} 
and $\Delta $ is a wavenumber increment that is $\ll |k_z|$ but
which remains finite as~$\nu \rightarrow 0$.
A solution to Equation~(\ref{eq:dEpmdt2})
is sought in which $E^\pm$ has a boundary
layer at some wavenumber $k_z = b(t)$.
For simplicity, the analysis is restricted to $k_z > 0$. The spectra
at negative $k_z$ values then follow from the fact that $E^+$
and~$E^-$ are even functions of~$k_z$.

It is useful to work with the dimensionless variables defined through the equations
\begin{equation}
k_z = s k_{z0} \qquad t = \tau\left[\frac{v_{\rm A}}{\pi
    k_{z0} (\delta v_0)^2 }\right]
 \qquad E^\pm = \frac{\tilde{E}^\pm(\delta
  v_0)^2}{k_{z0}} 
\qquad
\overline{ E^\pm} = \frac{\overline{ \tilde{E}^\pm}(\delta
  v_0)^2}{k_{z0}}  ,
\label{eq:dimensionless} 
\end{equation} 
where $k_{z0}$ is some characteristic wavenumber and $(\delta v_0)^2$
is the characteristic mean-square AW velocity fluctuation. The
analysis is restricted to the case in which
\begin{equation}
H \equiv \tilde{E}^+ = \tilde{E}^- .
\label{eq:defH} 
\end{equation} 
Equation~(\ref{eq:dEpmdt2}) then takes the form
\begin{equation}
\frac{\partial H}{\partial \tau} = s^2
H \frac{\partial }{\partial s}\left( s
  H\right) 
+ \epsilon \overline{ H} \frac{\partial^2
  H}{\partial s^2},
\label{eq:dEpmdtdim} 
\end{equation} 
where $\overline{H} = \overline{ \tilde{E}^+}$ and
\begin{equation}
\epsilon = \frac{\nu v_{\rm A}}{\pi k_{z0}^4} \ll 1.
\label{eq:defepsilon} 
\end{equation} 
Upon setting $H = H(x, \tau)$ in Equation~(\ref{eq:dEpmdtdim}), where
\begin{equation}
\epsilon x = s - \tilde{b} \qquad \tilde{b} = \frac{b}{k_{z0}},
\label{eq:defx} 
\end{equation} 
and discarding terms $\ll \epsilon^{-1}$, one obtains
\begin{equation}
0 = \left(\frac{d\tilde{b}}{d\tau} + \tilde{b}^3 H\right)\frac{\partial
  H}{\partial x} + \overline{ H} \frac{\partial^2
  H}{\partial x^2},
\label{eq:BL1} 
\end{equation} 
where $\overline{ H}$ is independent of~$x$.

As in Equations~(\ref{eq:Efb}) and (\ref{eq:trunc2}), it is assumed
that $H = 0 $ on the small-$k_z$ side of the boundary layer
at~$s = \tilde{b}$. The value of $H$ at dimensionless wavenumbers slightly larger
than~$\tilde{b}$ is denoted $H_>$.  The solution to
Equation~(\ref{eq:BL1}) must thus satisfy the boundary conditions
\begin{equation}
H \rightarrow \left\{ \begin{array}{ll}
0 & \mbox{ as $x \rightarrow - \infty$} \\
H_> & \mbox{ as $x\rightarrow \infty$}
\end{array},
\right.
\label{eq:Hbc} 
\end{equation} 
with $\overline{ H} = H_> /2 $.
Upon substituting 
\begin{equation}
H = \frac{H_>}{2}\left(1 +  \tanh\frac{x}{a}\right)
\label{eq:Hform} 
\end{equation} 
into Equation~(\ref{eq:BL1}), one finds that
\begin{equation}
0 = \left(\frac{d\tilde{b}}{d\tau}\right) \frac{H_>}{2a} \sech^2\left(\frac{x}{a}\right) + \frac{H_>^2\tilde{b}^3}{4a}\left(1 +
  \tanh\frac{x}{a}\right) \sech^2 \left(\frac{x}{a}\right) -
\frac{H_>^2}{2a^2} \sech^2 \left(\frac{x}{a}\right) \tanh\frac{x}{a}.
\label{eq:tanhsech} 
\end{equation} 
By separately equating the $\sech^2(x/a)$ terms and the $\tanh(x/a)
\sech^2(x/a)$ terms, one obtains
\begin{equation}
\frac{d\tilde{b}}{d\tau} = - \frac{H_> \tilde{b}^3 }{2}
\label{eq:dbdtau} 
\end{equation} 
and
\begin{equation}
a = \frac{2}{\tilde{b}^3}.
\label{eq:vala} 
\end{equation} 
Thus, the desired solution to Equation~(\ref{eq:BL1}) is Equation~(\ref{eq:Hform}) with
$a$ and $\tilde{b}$ given by Equations~(\ref{eq:dbdtau}) and (\ref{eq:vala}).
Equation~(\ref{eq:dbdtau}) can be rewritten in terms of the original
variables as
\begin{equation}
\frac{db}{dt} = - \frac{\pi b^3 E_>}{2v_{\rm A}},
\label{eq:dbdt} 
\end{equation} 
where $E_> = H_> (\delta v_0)^2/k_{z0}$, which is
equivalent to Equation~(\ref{eq:dbdt0}).

\section{Approximate Analytic Solution for Decaying, Cross-Helical Alfv\'en Waves in the Nonlinear Regime}
\label{ap:approx} 

In this appendix, a solution to
Equation~(\ref{eq:dEpmdt}) is sought in which 
\begin{equation}
E^+ = H(k_z - b) \sum_{n=1}^\infty c_n \left(\frac{k_z}{b}\right)^{-n}
+ H(b-k_z) \eta k_z^p,
\label{eq:Epseries} 
\end{equation}
\begin{equation}
E^- = H(k_z-b) \sum_{n=2}^\infty a_n \left(\frac{k_z}{b}\right)^{-n},
\label{eq:Emseries} 
\end{equation} 
and 
\begin{equation}
c_{n+1} \ll c_n \qquad a_{n+1} \ll a_n \qquad a_2 \ll c_1,
\label{eq:ca} 
\end{equation} 
where $H(x)$ is the Heaviside function, $a_n$,
$b$, and $c_n$ are functions of time, and $\eta$ and $p$ are constants, with
\begin{equation}
-1 < p < 1.
\label{eq:prange} 
\end{equation} 
Substituting
Equations~(\ref{eq:Epseries}) and (\ref{eq:Emseries})  into 
Equation~(\ref{eq:dEpmdt}) and making use of
Equation~(\ref{eq:ca}) leads to the following approximate solutions for
$b$, $c_1$, and~$a_2$:
\begin{equation}
b = \left[\frac{(p+2)(p+1)\pi \eta t  }{(1-p)v_{\rm A}} +
    \Lambda\right]^{-1/(p+2)},
\label{eq:b(t)} 
\end{equation} 
where $\Lambda$ is a constant of integration,
\begin{equation}
c_1 = \frac{(p+3)\eta b^p}{1-p},
\label{eq:c1} 
\end{equation} 
and
\begin{equation}
a_2 = \frac{(p+1)^2\eta b^p}{1-p}.
\label{eq:a2} 
\end{equation} 
Equations~(\ref{eq:b(t)}) through (\ref{eq:a2})  imply that the break
wavenumber $b(t)$ decreases in time, and that the ratios 
$E^+(b_+)/\eta b_+^p$ and $E^+(b_+)/E^-(b_+)$ remain constant, where
$b_+$ is a wavenumber infinitesimally larger than~$b$. For example,
if the infrared spectral index~$p$ is~$-1/2$, then 
\begin{equation}
\frac{E^+(b_+)}{E^-(b_+)} = \frac{p+3}{(p+1)^2} \rightarrow 10,
\label{eq:ratio1} 
\end{equation} 
and
\begin{equation}
\frac{E^+(b_+)}{\eta b_+^p} = \frac{p+3}{1-p} \rightarrow \frac{5}{3}.
\label{eq:ratio2} 
\end{equation} 
Equations~(\ref{eq:ratio1}) and (\ref{eq:ratio2}) are
shown as dotted lines in Figures~\ref{fig:num_wide} and~\ref{fig:num_narrow}.

\bibliography{articles} 

\bibliographystyle{jpp}

\end{document}